\documentclass[12pt]{article}
\usepackage[square,numbers,sort&compress]{natbib}
\usepackage{enumitem}
\usepackage{amsfonts}
\usepackage{amsmath}
\usepackage{algorithm}
\usepackage{algpseudocode}

\usepackage{soul}
\usepackage{graphicx}
\usepackage{subcaption}
\usepackage{fancyhdr}
\usepackage{amssymb}
\usepackage{float}
\usepackage{verbatim}
\usepackage{mathrsfs}

\usepackage[square,numbers,sort&compress]{natbib}
\usepackage{array,graphicx}
\usepackage{enumerate}
\usepackage{graphicx}
\usepackage{multirow}
\usepackage{leftidx}
\usepackage{setspace}
\usepackage{tikz}
\usetikzlibrary{shapes.geometric, arrows}
\usepackage[titletoc,toc,title]{appendix}
\usepackage{amsmath}
\usepackage{graphicx}

\usepackage{colortbl}
\usepackage{booktabs}
\usepackage{booktabs}
\usepackage{multirow}

\usepackage{setspace}
\newcommand{\R}{\mathbb{R}}
\newcommand{\E}{\mathbb{E}}
\newcommand{\diff}{\mathrm{d}}

\newcommand{\examplenum}{three }


\begin{document}
\title{A deep learning-driven iterative scheme for high-dimensional HJB equations in portfolio selection with exogenous and endogenous costs}
\author{Dong Yan$^a${\thanks{Corresponding author: dyan@uibe.edu.cn}, Nanyi Zhang$^b$, Junyi Guo$^c$}\\
\small\it a. School of Statistics, University of International Business and Economics, Beijing, China.\\
\small\it b. School of Mathematical Sciences, Peking University, Beijing, China.\\
{\small\it c. School of Mathematical Sciences, Nankai University, Tianjin, China}\\
}
\date{}
\maketitle

\begin{abstract}
In this paper, we first conduct a study of the portfolio selection problem, incorporating both exogenous (proportional) and endogenous (resulting from liquidity risk, characterized by a stochastic process) transaction costs through the utility-based approach. We also consider the intrinsic relationship between these two types of costs. To address the associated nonlinear two-dimensional Hamilton-Jacobi-Bellman (HJB) equation, we propose an innovative deep learning-driven policy iteration scheme with three key advantages: i) it has the potential to address the curse of dimensionality; ii) it is adaptable to problems involving high-dimensional control spaces; iii) it eliminates truncation errors. The numerical analysis of the proposed scheme, including convergence analysis in a general setting, is also discussed. To illustrate the impact of these two types of transaction costs on portfolio choice, we conduct through numerical experiments using \examplenum typical utility functions.

\end{abstract}


\section{Introduction}
Portfolio selection, which aims to optimally allocate an investor's wealth between risky assets and risk-free bonds, is well-known as a classical topic in mathematical finance. It has attracted significant attention from both academic scholars and financial practitioners since the pioneering contributions of Markowitz's single-period mean-variance analysis \cite{Markowitz1952} and Merton's continuous-time expected utility maximization \cite{Merton1971} were introduced. So far, the portfolio selection problem has been thoroughly studied in the context of a complete market. However, many researchers have pointed out that the market is not perfectly frictionless, as every transaction in stocks incurs transaction costs. These costs can be categorized into two main types: exogenous transaction costs and endogenous transaction costs.

Exogenous transaction costs refer to fees that are external to the transaction itself and generally include various taxes and expenses in the real market. In practice, it is not easy to find a realistic representation of these transaction fees. Under the framework of Markowitz \cite{Markowitz1952}, some researchers analyzed the impact on portfolio selection of fixed transaction costs \cite{Patel1982}, quadratic transaction costs \cite{Peng2008}, and a specific form of minimum transaction costs \cite{Baule2010}. Additionally, many other studies \cite{Dai2010, Wang2013, Dybvig2020, Pun2022, Mei2023} assumed that the transaction costs are proportional to the monetary value of the transaction, with a constant transaction cost rate, which are closer to practice. However, most of these papers cannot be extended to deal with the multi-period case, as the model will become time inconsistent due to the failure of the Bellman optimality principle. Such a limitation implies that dynamic programming cannot be easily applied, and thus the optimal strategy for trading is hard to be definitively determined. To address this problem, other researchers \cite{Davis1990, Chellathurai2005, Choi2007, Dai2008} studied the problem via the utility maximization theory, as Merton's continuous-time formulation \cite{Merton1971} offers a powerful analytical framework, enabling the extension of the one-period mean-variance portfolio model to dynamic scenarios. From a mathematical point of view, it is equivalent to solving a high-dimensional HJB equation, which poses challenges both analytically and numerically.

In contrast to exogenous transaction costs, endogenous transaction costs are those that are internal to the transaction and are directly related to the actions of the transacting parties. One of the most typical endogenous transaction costs is liquidity cost, resulting from liquidity risk, which is almost ubiquitous in the real market. Given the noticeable impact of liquidity risk on stock prices as shown in \cite{Feng2014}, liquidity risk should be incorporated into the pricing of financial derivatives and also the process of optimal asset allocation. Then, how to effectively model liquidity risks is an intriguing problem that has attracted researchers to study the features of liquidity risks both financially and empirically. Some works utilized different measures of individual stock illiquidity in the real market, such as the illiquidity ratio and the bid-ask spread \cite{Gonzalez2011, Abensur2022, Niu2012}. Other works proposed various market impact models of liquidity within a Value at Risk (VaR) framework \cite{Al2021} or modeled the market impact costs due to liquidity risks explicitly as a function \cite{Ha2020, Ly2007, Caccioli2016}. Meanwhile, another group of researchers depicted liquidity risks using a stochastic process \cite{Feng2014, Feng2016, Pasricha2022, He2024}. Specifically, Feng et al. \cite{Feng2014} assumed that liquidity risk follows a stochastic mean-reverting process and applied a market liquidity-related discounting factor to calculate stock prices affected by liquidity risk. In their subsequent work \cite{Feng2016}, they further empirically demonstrated the good performance of incorporating such a formulation of liquidity risk into European option pricing problems, compared to the case without liquidity risk. Recently, Pasricha et al. \cite{Pasricha2022} took a step further by incorporating a more general correlation structure among different Brownian motions, generalizing Feng et al.'s model \cite{Feng2014}.

Driven by the significance of exogenous and endogenous transaction costs, this paper analyzes how these factors influence an investor's portfolio decisions under the utility-based approach, considering liquidity risks and transaction costs associated with asset trading. Based on the works of Pasricha et al. \cite{Pasricha2022}, we adopt a mean-reverting Ornstein–Uhlenbeck (OU) process for stochastic liquidity. Furthermore, we adjust the long-run market liquidity level, taking into account the intrinsic connection between exogenous and endogenous transaction costs.

Mathematically, the core of such a portfolio selection problem is to solve a two-dimensional Hamilton-Jacobi-Bellman (HJB) equation. Since the problem is highly nonlinear, we adopt the method based on physics-informed neural networks (PINNs) \cite{Raissi2019} and propose an innovative deep learning-driven policy iteration scheme to effectively obtain numerical solutions. Different from the classic policy iteration schemes \cite{Forsyth2007, Alla2015, Kerimkulov2020} that solve the associated partial differential equation (PDE) via traditional mesh-based methods, our proposed numerical scheme has three advantages: i) It is mesh-free, which potentially enables it to address the curse of dimensionality \cite{Grohs2023}, and is therefore applicable to models that incorporate various market frictions, such as non-constant volatility and other market risks; ii) The control in the HJB equation is approximated by a non-discretized two-layer neural network, allowing our scheme to be extended effectively to solve problems with high-dimensional control spaces; iii) Derivatives are computed using automatic differentiation \cite{Baydin2018}, thereby eliminating truncation errors.

The remainder of this paper is organized as follows. In Section 2, we present the details of the formulation of the HJB equation for the portfolio maximization problem, incorporating stochastic liquidity risk and proportional transaction costs. The proposed deep learning-driven policy iteration scheme, along with its numerical analysis, is presented in Section 3. Then we present three examples to analyze the impact of both exogenous and endogenous transaction costs on portfolio decisions in Section 4. Concluding remarks are provided in the last section.

\section{Formulation of the model}

In this section, we formulate a dynamic portfolio selection model, where both proportional transaction costs when trading stocks to rebalance the portfolio and stochastic liquidity risks are considered. Consider a financial market consisting of two assets. One is the risk-free bond $x(t)$ whose dynamics is $dx(t)=rx(t)dt$ with $r \geq 0$ being the risk-free interest rate. Another one is the risky stock $S(t)$, and followed by the detailed derivations in \cite{Pasricha2022}, the stock price $S$ and the liquidity risk $L$ can be characterized as

\begin{equation}\label{eq:dynamics-stock}
\begin{cases}
        d S_{t}= \mu S_t d t+\beta L_tS_t d B^\gamma_t+\sigma_SS_t d B^S_t,\\
         d L_{t}=\alpha(\theta(L_t)-L_t) d t+\sigma_L d B^L_t,
\end{cases}
\end{equation}

where $B^\gamma$, $B^S$, $B^L$ are correlated standard Brownian motions specified as
\begin{equation}
\begin{cases}
   d B^\gamma_t d B^S_t=\rho_1dt,  \\
         d B_t^L d B^S_t=\rho_2 d t,\\
        d B^\gamma_t d B^L_t=\rho_3 d t.
\end{cases}\label{eqcorr}
\end{equation}
$\sigma_S$ is the constant volatility of the asset price, and the strictly positive parameter $\beta$ measures the sensitivity to the level of market liquidity of the asset price. The liquidity risk $L$ follows a mean-reverting process with the mean-reversion speed $\alpha$, the mean-reversion level $\theta$ and the volatility of market liquidity $\sigma_L$.

When transaction costs are taken into consideration, two important factors need to be considered in dynamic portfolio optimization. First, the portfolio cannot be continuously rebalanced, as it would lead to abnormally large trading costs. Therefore, the portfolio is hedged discretely in a non-infinitesimal fixed time step, denoted as $\delta t$. Second, the intrinsic link between transaction fees and liquidity risks should not be eliminated. In theory, an increase in the transaction costs rate will enlarge the market's illiquidity, as investors become more reluctant to trade due to the reduced profit margins. To capture this effect, we assume that the mean-reversion of the illiquidity level is positively correlated with the transaction costs rate. Specifically, in Equation (\ref{eq:dynamics-stock}), we assume that the mean-reversion level of liquidity risks is influenced not only by the current illiquidity level (denoted as $\bar{\theta}$, which excludes the effects of transaction costs), but also by the transaction costs rate and a transaction costs sensitivity coefficient $\lambda$:
\begin{equation}
\theta(L)=\bar{\theta}+\kappa\cdot  \lambda \cdot g(L).
\label{theta}
\end{equation}
Since the effects of transaction costs on illiquidity are expected to decrease as the illiquidity level increases, $g(L)$ should be selected as a concave function. Here, we use the power function as an illustrative example, where $g(L)=L^\zeta$, with $\zeta\in (0, 1)$.

Now we consider an investor with an initial endowment $W_0$ invests a fraction of $\omega(t)\in [0, 1]$ of his or her total wealth $W(t)$ on the underlying assets at time $t$, then $1-\omega(t)$ is the fraction of the wealth left in the form of risk-free asset. In order to limit the accumulated trading costs, we assume that the portfolio is hedged in a non-infinitesimal time step denoted by $\delta t$, and such exogenous transaction cost is proportional to the value of traded stocks during these fixed regular intervals. Thus, the net change of the investor's wealth in one time step can be expressed as

\begin{equation}\label{eq:dynamics-wealth}
\delta W=(r W + (\mu -r)\omega W )\delta t+\omega \beta L W \delta B^\gamma + \omega\sigma_S W \delta W^S-\kappa S |\nu|.
\end{equation}

Now we need to obtain the explicit solution of $\nu$, i.e. the number of traded stocks during $\delta t$. Since the number of stocks held at time $t$ is $\frac{\omega(t) W(t)}{S(t)}$, the number of traded stock from time $t$ to $t+\delta t$ can be written as
\begin{equation*}
 \nu=\delta \bigg(\frac{\omega W}{S}\bigg).
\end{equation*}
Then, we apply It$\hat{\text{o}}$'s lemma to $\nu$, and keep the terms of $O(\sqrt{\delta t})$,
\begin{equation}
\nu=\frac{(\omega-1)\omega W}{S}\cdot \bigg[\frac{\beta L \delta B^\gamma+\sigma_S \delta B^S}{1+\kappa\cdot \text{sign}(\nu)\omega}\bigg].
\end{equation}
We cannot predict the exact number of traded stocks beforehand, but the expected transaction costs can be computed in a time step as follows
\begin{equation}
\mathbb{E} \{ \kappa S |\nu| \}=\frac{\kappa}{1+\kappa\cdot \text{sign}(\nu)\omega}\cdot |\omega-1|\omega W\cdot \mathbb{E}\bigg\{ \bigg| \beta L \delta B^\gamma+\sigma_S \delta B^S \bigg| \bigg\}.
\label{E_init}
\end{equation}
Noting the transaction costs rate $\kappa<1$ and the fraction $\omega \in [0, 1]$, $\kappa^2 \omega \ll 1$, then
\begin{equation}
\frac{\kappa}{1+\kappa \cdot \text{sign}(\nu)\omega}\approx\kappa \pm \kappa^2 \omega \approx \kappa.
\end{equation}
To compute the last term in Eq. (\ref{E_init}), since $\delta B^\gamma$ and $\delta B^S$ are correlated with value $\rho_1$, we can write 
\begin{align*}
&\delta B^\gamma=\sqrt{\delta t} Z_1,\\
&\delta B^S=\rho_1\sqrt{\delta t} Z_1+\sqrt{1-\rho_1^2}\sqrt{\delta t} Z_2,
\end{align*}
where $Z_1, Z_2 \sim \mathscr{N}(0, 1)$, and thus
\begin{equation}
\mathbb{E}\bigg\{ \bigg| \beta L \delta B^\gamma+\sigma_S \delta B^S \bigg| \bigg\}=\sqrt{\frac{2}{\pi}}\cdot\sqrt{(\beta L+\sigma_S \rho_1)^2+(1-\rho_1^2)\sigma_S^2}\cdot  \sqrt{\delta t}
\end{equation}
Then the expected transaction costs in one time step can be approximated as
\begin{equation}
\mathbb{E} \{ \kappa S |\nu| \}=\sqrt{\frac{2}{\pi \delta t}}\kappa (1-\omega)\omega W \sqrt{(\beta L+\sigma_S \rho_1)^2+(1-\rho_1^2)\sigma_S^2}\cdot \delta t. 
\end{equation}
Therefore, with both exogenous and endogenous transaction costs being taken into consideration, the investors' wealth process should satisfy the following dynamics

\begin{equation}
\label{eq:dynamics-wealth-v2}
\delta W=(r W + (\mu -r)\omega W )\delta t+\omega \beta L W \delta B^\gamma +\omega\sigma_S  W \delta W^S-\sqrt{\frac{2}{\pi \delta t}}\kappa (1-\omega)\omega W \sqrt{(\beta L+\sigma_S \rho_1)^2+(1-\rho_1^2)\sigma_S^2}\cdot \delta t.
\end{equation}

We are now ready to formulate a utility maximization model for the investor who invests wealth in bond and stock. The investor aims to maximize the expected utility of his or her terminal wealth at time $T$ by using an admissible trading strategy $\mathscr{A}$, i.e. by adjusting the fractions invested in stocks and bonds. Then, the value function $Q$ can be stated as
\begin{equation}
Q(W, L, t)=\max_{\omega \in \mathscr{A}} \mathbb{E}_t \bigg\{ \mathscr{U}(W(T)) \bigg| W(t)=W, L(t)=L \bigg\},
\end{equation}
where $ \mathscr{U}(\cdot)$ is the investor's utility function. In this work, two types of utilities including CRRA(constant relative risk aversion) and CARA(constant absolute risk aversion) are applied to represent different investors' risk preference. 

With the dynamics of state variables specified in Eqs. (\ref{eq:dynamics-stock}) and (\ref{eq:dynamics-wealth-v2}), the HJB equation is derived as
\begin{equation}
\label{Q}
\max_{\omega \in [0, 1]} \bigg\{\mathscr{L}^{\omega} Q(W,L,t) \bigg\}=0, \quad \forall (W,L,t) \in \Omega_T,
\end{equation}
where $\Omega_T = \R_+ \times \R_+ \times [0,T]$ and the operator $\mathscr{L}^\omega,\ $ is given by
\begin{align}
\label{L}
\mathscr{L}^\omega &=
\frac{\partial }{\partial t}+\bigg(r W + (\mu -r)\omega W -\sqrt{\frac{2}{\pi \delta t}}\kappa \omega (1-\omega) W \sqrt{(\beta L+\sigma_S \rho_1)^2+(1-\rho_1^2)\sigma_S^2}\bigg)\frac{\partial }{\partial W}\\ \nonumber
&+\frac{1}{2}\bigg(\beta^2 L^2+\sigma_S^2+2\rho_1 \sigma_S \beta L\bigg)\omega^2 W^2 \frac{\partial ^2 }{\partial W^2}+\alpha (\theta-L) \frac{\partial }{\partial L}+\frac{1}{2}\sigma_L^2 \frac{\partial ^2 }{\partial L^2}+\bigg(\rho_2 \sigma_S+\rho_3 \beta L\bigg)\sigma_L \omega W \frac{\partial ^2 }{\partial W\partial L}, \nonumber
\end{align}
with the terminal condition $Q(W, L, T)=\mathscr{U}(W)$.

\section{Solution for the HJB equation based on the deep learning-driven policy iteration scheme}
In this section, we propose a deep learning-driven policy iteration scheme to solve Eq. (\ref{Q}) numerically, which is proven to be monotonically increasing and conditionally convergent. To validate our numerical scheme, we compare the numerical solutions obtained with zero liquidity risk and transaction costs with the results from the analytical solution of Merton's problem under a power utility function.

\subsection{The classic policy iteration scheme}
Obviously, the operator $\mathscr{L}^\omega$ specified in Eq. (\ref{L}) is quadratic in $\omega$, then the optimal policy of the corresponding maximum function can be expressed explicitly with conditions. However, the quotient of partial derivatives may cause computational problems. To find optimal policies for the portfolio selection problem, the policy iteration scheme is considered to be the most efficient method. Remarkably, the policy iteration scheme often converges to the solution within a surprisingly limited number of iterations. Since our proposed numerical scheme is based on the policy iteration scheme, we recall the fundamentals of the classic policy iteration scheme \cite{Howard1960} beforehand.

We assume that the estimated policy at each iteration is denoted by $\omega_k \in [0, 1]^{\Omega_T}, k \in \{0, 1, \cdots, N\}$. With a given initial policy $\omega_0$, the optimal policy $\omega^*$ and the numerical result of Eq. (\ref{Q}) can be obtained through a process of iterative evaluation and improvement until certain termination criterion is met. In the step of \textbf{policy evaluation (PE)}, an approximate value function at $k$-th iteration, denoted by $Q_k$, can be obtained by solving the equation $\mathscr{L}^{\omega_{k-1}}Q_k=0$, subject to the terminal condition. Then in order to yield a better policy, namely through the step of \textbf{policy improvement (PI)}, one needs to find $\omega\in [0, 1]$ that maximizes the values of $\mathscr{L}^\omega Q_k(W,L,t)$ for each $(W,L,t)$ belonging to $\Omega_T$. Then we can obtain the improved policy which is denoted by $\omega_k$. Further analysis will demonstrate that the induced value function $Q_k$ is monotonically increasing with each iteration, and a sequence of monotonically improving policies and approximate solutions are obtained:

\begin{equation*}
\omega_{0}\xrightarrow[ ]{ \ \text{PE} \ }Q_1 \xrightarrow[ ]{ \ \text{PI} \ }\omega_{1}\xrightarrow[ ]{ \ \text{PE} \ }Q_2 \xrightarrow[ ]{ \ \text{PI} \  }\omega_{2}\xrightarrow[ ]{ \ \text{PE} \ }\cdots\xrightarrow[ ]{ \ \text{PI} \ }\omega^{\ast}\xrightarrow[ ]{ \ \text{PE} \ }Q^*
\end{equation*}

\subsection{The deep learning-driven policy iteration scheme}
It should be pointed out that the essence of the policy iteration scheme lies in solving the associated PDE given a policy (also known as the process of policy evaluation). Many researchers engaged studies by employing the finite difference method and demonstrated good convergence performance for their numerical schemes \cite{Forsyth2007, Greif2017, Kerimkulov2020}. However, such traditional mesh-based methods may encounter truncation errors, and the computational efficiency will be greatly lowered with the addition of an extra dimension in space \cite{Lu2021}. In particular, when applying the implicit finite difference method to high-dimensional problems, the cross derivative terms significantly complicate the computation of the inverse matrix.

An alternative approach to solving PDEs utilizes deep learning, which has emerged as a powerful tool in recent years, achieving remarkable success across a broad spectrum of applications. In this work, we adopt the method derived from PINNs \cite{Raissi2019}, chosen for our problem owing to its two notable advantages: i) the PINNs algorithm is characterized by its simplicity, making it adaptable to various types of PDEs; ii) it embeds the PDE directly into the loss function of the neural networks using automatic differentiation, offering a mesh-free approach that can potentially address the curse of dimensionality.

In our algorithm, the value function and the optimal policy are approximated by corresponding well-constructed neural networks. This is because, according to the Universal Approximation Theorem, under mild assumptions about the activation function, a two-layer neural network can closely approximate any continuous function defined on compact subsets \cite{Funahashi1989}. 

We define the value network $Q_\phi: \Omega_T \to \R$ and the control network $\omega_\psi: \Omega_T \to [0, 1]$ \footnote{The control network can be extended to accept higher-dimensional input spaces and produce higher-dimensional output spaces as well, accommodating more complex control scenarios.} as two-layer neural networks to approximate $Q^*$ and $\omega^*$ in Eq. (\ref{Q}) respectively, where
\begin{align}
Q_\phi &= f^{Q}_2 \circ \mathbf{tanh} \circ f_1^Q,\\\nonumber
\omega_\psi &= \mathrm{sigmoid} \circ f^{\omega}_2 \circ \mathbf{tanh} \circ f_1^\omega,
\end{align}
with parameters specified in Appendix \ref{appendixA}. Note that $\phi= [W_1^Q, b_1^Q, W_2^Q, b_2^Q]$ and $\psi= [W_1^\omega, b_1^\omega, W_2^\omega, b_2^\omega]$ represent the trainable parameters of the networks, comprising the weights and biases at each layer, respectively.

During the step of policy evaluation, solving the associated PDE is achieved by minimizing the loss function, which is defined as the weighted summation of the PDE residuals and the terminal condition residuals. Specifically, the loss function during the $k$-th iteration is given by
\begin{equation}\label{eq:loss-func}
\mathcal{L}^{(k)}(\phi) = \E_{W, L, t} \left[ \left(\mathscr{L}^{\omega_{\psi^{(k-1)}}}Q_\phi(W, L, t) \right)^2\right] + \E_{W,L}\left[\left| Q_\phi(W,  L, T) - \mathscr{U}(W) \right|^2 \right].
\end{equation}
Note that $\E_{W, L, t}$ denotes the expectation with respect to the variables $(W, L, t)$. Instead of constructing these variables by following a uniform distribution within the pre-specified domain $\Omega_T$, we adopt the idea proposed by Lu et al. \cite{Lu2021}, which involves utilizing a non-uniform distribution for the expectation. This is achieved by increasing the sampling density in regions with higher residuals, as empirical evidence supports that such a construction results in improved outcomes. In addition, the partial derivatives appears in \eqref{eq:loss-func} are calculated via automatic differentiation.

By minimizing this loss function, the updated trainable parameters of the value network at the k-th iteration, denoted $\phi^{(k)}$, are determined by the following optimization problem:
\begin{equation*}
\phi^{(k)} = \mathop{\mathrm{argmin}}_{\phi} \mathcal{L}^{(k)}(\phi).
\end{equation*}
After closely aligning the estimated value function with the actual dynamics dictated by the PDE and terminal conditions, we are ready to update the trainable parameters of the control network by maximizing the following equation using the newly updated value network parameters:
\begin{equation*}
\psi^{(k)} = \mathop{\mathrm{argmax}}_\psi \mathbb{E}_{W, L, t}\left[\mathscr{L}^{\omega_\psi}Q_{\phi^{(k)}}(W, L, t) \right].
\end{equation*}
This maximization step can be viewed as an empirical implementation of the policy improvement procedure, aiming to refine the current policy based on the improved estimations from the value function. As a result, the numerical scheme of Eq. (\ref{Q}) can be easily deduced as follows.
\begin{algorithm}[H]
\caption{The deep learning-driven policy iteration scheme}\label{algo:API}
\begin{algorithmic}[]
\State Given initial values of trainable parameters of networks $\phi^{(0)}$, $\psi^{(0)}$
\State Construct value network and control network
\For{$k = 1,\ldots, N$}
\State Conduct the step of \textbf{policy evaluation} by calculating
$$
\phi^{(k)} = \mathop{\mathrm{argmin}}_{\phi}\E_{W,L,t} \left[ \left(
\mathscr{L}^{\omega_{\psi^{(k-1)}}}Q_{\phi}(W,L, t) \right)^2\right] + \E_{W,L}\left[\left|Q_\phi(W,L,0) - \mathscr{U}(W) \right|^2 \right].
$$
\If{the difference between $Q_{\phi^{(k)}}$ and $Q_{\phi^{(k-1)}}$ is small}
\State \textbf{break} 
    \EndIf
\State Conduct the step of \textbf{policy improvement} by calculating
\begin{equation*}
\psi^{(k)} = \mathop{\mathrm{argmax}}_\psi \mathbb{E}_{W,L,t}\left[ \mathscr{L}^{\omega_\psi}Q_{\phi^{(k)}}(W,L,t)\right].
\end{equation*}
\EndFor
\State \textbf{return} $Q_{\phi^{(k)}}, \omega_{\psi^{(k)}}.$
\end{algorithmic}
\end{algorithm}

\subsection{Convergence Analysis} 
In this section, we perform convergence analysis of our proposed deep learning-driven policy iteration scheme in a general setting. Consider a stochastic control problem where the dynamics of the state variables $X_s$ are modeled by a controlled $\R^d$-valued It$\hat{\text{o}}$ diffusion:
\begin{equation}\label{eq:general-sde}
\diff X_s = b(t, X_s, u_s) \diff s + \sigma(s, X_s, u_s) \diff W_s, \ s\in[t, T], \ X_t=x,
\end{equation}
with the drift and diffusion coefficients $b, \sigma: \R^+ \times \R^d \times U \to \R^d, \R^{d \times m}$ respectively, and $U$ represents the control space. $W_s$ denotes a $d$-dimensional Brownian motion defined on a filtered probability space $(\Omega, \mathscr{F}, (\mathscr{F}_t)_{t\geq 0}, \mathbb{P})$.

By utilizing the admissible trading strategies $\mathcal{A}_s$, the investor aims to maximize the expected utility of their intermediate returns $f$ over the investment period $[t, T]$, combined with the terminal return $g$ at time $T$. Let $X^{t,x,u}$ to represent the solution of \eqref{eq:general-sde} starting from state $x$ at time $t$ under control $u$. We then consider the following weak formulation of a general optimal control problem:
\begin{equation}
\label{VF2}
V(t,x) = \sup_{{u \in \mathcal{A}_s} } \E \left[ \int_t^T f\left(s, X_s^{t,x,u}, u(s, X_s^{t,x,u})\right) \diff s + g(X_T^{t,x,u})\right].
\end{equation}

It can be easily derived that the value function satisfies the following HJB equation: 
\begin{equation}\label{eq:HJB-general}
\partial_t v + \displaystyle\sup_{u \in U} \left\{ f(t,x,u) +  b(t,x,u) \cdot \nabla v + \frac{1}{2} \mathrm{tr}[\sigma\sigma^T(t,x,u) \nabla^2 v]\right\}= 0,
\end{equation}
with the terminal condition $v(T, x) = g(x)$.

For the convenience of calculations, we define $H$, $H^\pi$ and $f^\pi$ as:
 \begin{equation}
 \begin{cases}
H(t,x,d,h,u) = f(t,x,u) + b(t,x,u) \cdot d + \frac{1}{2} \mathrm{tr}[\sigma \sigma^T(t,x,u) h], \\
H^\pi(t,x,v) = \partial_t v + H(t,x, \nabla v(t,x), \nabla^2v(t,x), \pi(t,x)), \\
f^\pi(t,x) = f(t,x, \pi(t,x)).
\end{cases}
\end{equation}

Through our proposed numerical method, the value function $V$ and the optimal control $u$ in Eq. (\ref{VF2}) are approximated by the value network $V_\phi$ and the control network $u_\psi$, respectively. According to Algorithm \ref{algo:API}, with the given initial values of the trainable parameters of the networks, the value function can be computed by iteratively conducting the corresponding steps of policy evaluation and policy improvement until the difference between the values of the value function at the $k$-th and $(k+1)$-th iterations is small enough. To simplify notation, at the $k$-th iteration, we denote:
\begin{equation*}
V^{\pi_k} := V_{\phi^{(k)}} \ \text{and} \ \pi_{k} := u_{\psi^{(k-1)}}.
\end{equation*}

Let $e_k$ denote the error in solving the PDE in the k-th policy evaluation procedure. Then we have
\begin{equation*}
H^{\pi_k}(t,x,V^{\pi_{k}}) = e_k(t,x),
\end{equation*}
and 
\begin{equation*}
H^{\pi_{k+1}}(t,x,V^{\pi_{k}}) = \sup_{u \in U} H^{(t,x)\mapsto u}(t,x,V^{\pi_{k}}) \geq H^{\pi_{k}}(t,x,V^{\pi_{k}}) = e_k(t,x).
\end{equation*}

Under the feasible control $\pi_k$, we assume that the corresponding value function $V^{\pi_k}$ and intermediate returns $f^{\pi_k}$ grow quadratically with respect to $x$. We introduce a sequence of stopping times $\tau_n$ defined by 
$$
\tau_n = T \wedge \inf \{s > t: |X_s^{t, x, \pi_{k+1}} - x| \geq n\},
$$
to restrict the controlled process $X^{t,x,\pi_{k+1}}$ to a bounded region.

By applying Itô’s formula to $V^{\pi_{k}}\bigl(s,X_s^{t,x,\pi_{k+1}}\bigr)$, we obtain:
\begin{align*}
V^{\pi_{k}}(t, x) =& \E\left[V^{\pi_{k}}(\tau_n, X_{\tau_n}^{t, x,\pi_{k+1}})\right] - \E\left[\int_t^{\tau_n} (\partial_t + \mathcal{A}^{\pi_{k+1}})V^{\pi_{k}}(s,X_s^{t, x,\pi_{k+1}}) \diff s \right] \\
 &- \E \left[\int_t^{\tau_n} \partial_x V^{\pi_k}(s, X_s^{t,x,\pi_{k+1}})^T \sigma(s, X_s^{t,x,\pi_{k+1}}) \diff W_s\right]. 
\end{align*}

Since the last integrand term is bounded, the stochastic integral is a martingale, and thus,
\begin{align*}
V^{\pi_{k}}(t, x) =& \E\left[V^{\pi_{k}}(\tau_n, X_{\tau_n}^{t, x,\pi_{k+1}})\right] - \E\left[\int_t^{\tau_n} (\partial_t + \mathcal{A}^{\pi_{k+1}})V^{\pi_{k}}(s,X_s^{t, x,\pi_{k+1}}) \diff s \right] \\
=& \E\left[V^{\pi_{k}}(\tau_n, X_{\tau_n}^{t, x,\pi_{k+1}})\right] + \E\left[\int_t^{\tau_n} f^{\pi_{k+1}}(s,X_s^{t, x,\pi_{k+1}}) \diff s \right] \\ 
&- \E\left[\int_t^{\tau_n} H^{\pi_{k+1}}(s,X_s^{t, x,\pi_{k+1}},V^{\pi_{k}}) \diff s \right] \\
\leq& \E\left[V^{\pi_{k}}(\tau_n, X_{\tau_n}^{t, x,\pi_{k+1}})\right] + \E\left[\int_t^{\tau_n} f^{\pi_{k+1}}(s,X_s^{t, x,\pi_{k+1}}) \diff s \right] + (T-t) \Vert e_k \Vert_\infty \\
=& \E\left[V^{\pi_{k}}(\tau_n, X_{\tau_n}^{t, x,\pi_{k+1}})\right] + \E\left[\int_t^{\tau_n} f(s,X_s^{t, x,\pi_{k+1}}, \pi_{k+1}(s, X_s^{t, x,\pi_{k+1}})) \diff s \right] + (T - t) \Vert e_k \Vert_\infty.
\end{align*}

Since $V^{\pi_k}$ and $f^{\pi_k}$ are assumed to have quadratic growth in $x$, the sum of the absolute values of the first two key terms is bounded and integrable, which can be written as
\begin{align*}
& \left | V^{\pi_{k}}(\tau_n, X_{\tau_n}^{t, x,\pi_{k+1}}) \right | + \left |\int_t^{\tau_n} f(s,X_s^{t, x,\pi_{k+1}}, \pi_{k+1}(s, X_s^{t, x,\pi_{k+1}})) \diff s \right | \\
&\hspace{1.5cm} \leq C_1 \left( 1 + \sup_{t \leq s \leq T} |X_s^{t,x,\pi_{k+1}}|^2\right) \in L^1(\Omega).
\end{align*}

Then applying the dominated convergence theorem yields
\begin{align*}
V^{\pi_{k}}(t, x) &\leq \E\left[\int_t^T f\left(s,X_s^{t, x,\pi_{k+1}}, \pi_{k+1}(s,X_s^{t, x,\pi_{k+1}})\right) \diff s + V^{\pi_{k}}(T, X_T^{t,x,\pi_{k+1}}) \right] + (T-t) \Vert e_k \Vert_\infty\\
&= \E\left[\int_t^T f\left(s,X_s^{t, x,\pi_{k+1}}, \pi_{k+1}(s,X_s^{t, x,\pi_{k+1}})\right) \diff s + g(X_T^{t,x,\pi_{k+1}}) \right] + (T-t) \Vert e_k \Vert_\infty \\
&= V^{\pi_{k+1}}(t, x) + (T-t) \Vert e_k \Vert_\infty.
\end{align*}

In practice, the error $e_k$ resulting from the solution of the PDE is small enough to be negligible, implying that the iteration algorithm is increasing monotonically.

Moreover, the convergence of our numerical scheme can be proved under the following restrictive assumption:
\begin{equation}
\label{assump}
H^{\pi_{k+1}}(t,x,V^{\pi_k}) \to 0 \ \text{uniformly},
\end{equation}
which was first proposed by Jacka and Mijatovi{\'c} \cite{Jacka2017}. 


\subsection{Validation of numerical scheme}
Since an analytical solution for the HJB equation (\ref{Q}) is not attainable, we trace back to the case of Merton's optimal allocation problem \cite{Merton1971} to validate our formulations. In this case, the parameters corresponding to liquidity risk and transaction costs in Eq. (\ref{Q}) are set to zero. Firstly, we compare the numerical results to the analytical solution of Merton's problem under a power utility function, $\mathscr{U}(W)=\frac{W^\gamma}{\gamma}$, with the parameters $\mu=0.05$, $r=0.02$, $\sigma_S=0.4$, $T=1$ and $\gamma=0.5$. Additionally, the experimental convergence rates are also discussed in this subsection to verify the convergence of our numerical scheme.

\begin{figure}[H]
    \centering
    \includegraphics[width=0.9\linewidth]{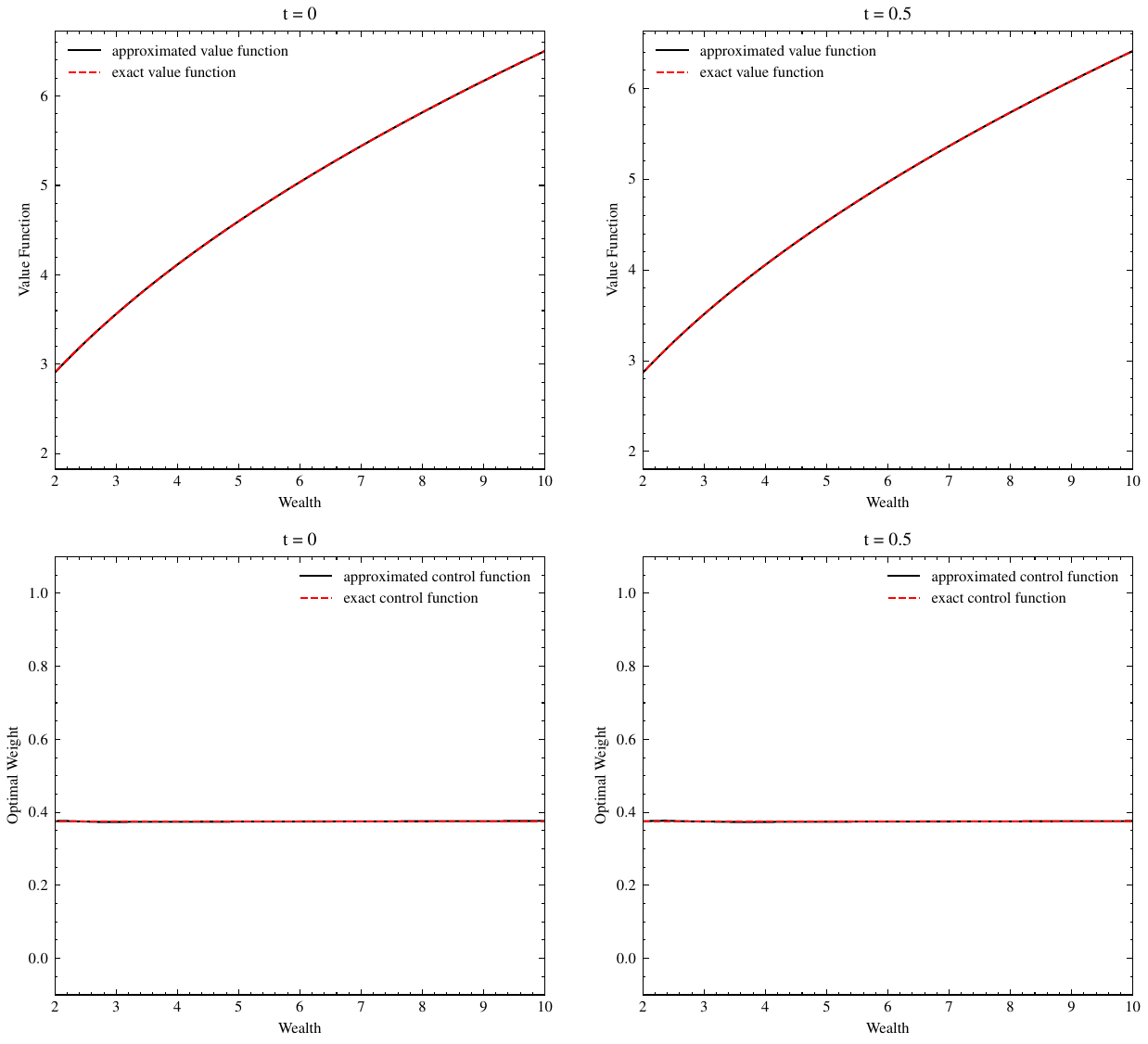}
    \caption{Validation of numerical scheme with zero liquidity risk and zero transaction costs}
    \label{fig:merton-compare}
\end{figure}

In Figure \ref{fig:merton-compare}, we compare both the values of the value function and the optimal policy at two time points: the initial time and the midpoint of expiry. The comparison shows that our algorithm yields a highly accurate estimation, closely matching the analytical solution. The optimal policy is a constant relative to an investor's wealth, which is theoretically consistent with the conclusion derived for Merton's problem, where $\omega^*=\frac{\mu-r}{(1-\gamma)\sigma^2}=0.375$.

\begin{figure}[H]
    \centering
    \includegraphics[width=0.5\linewidth]{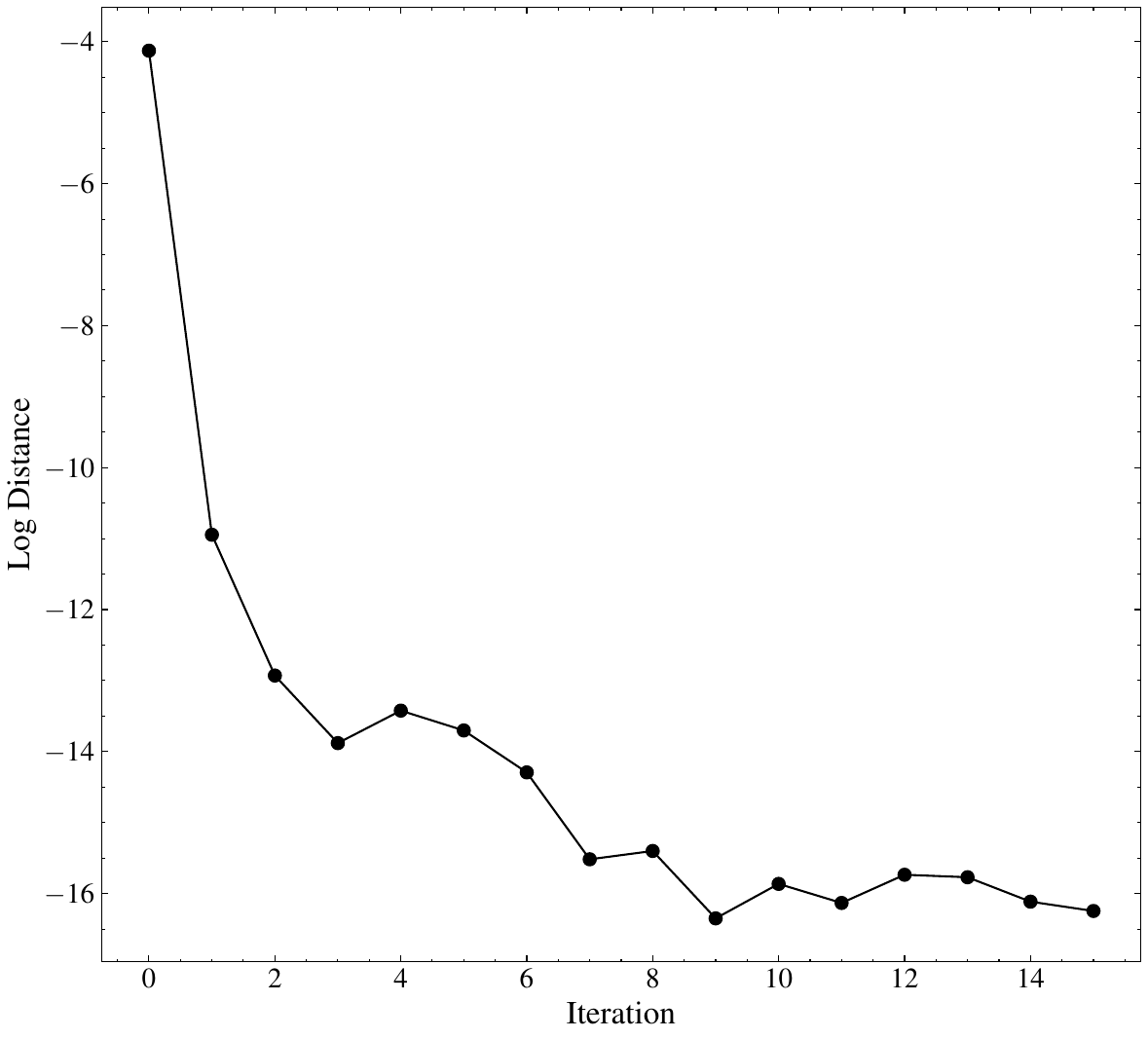}
    \caption{Experimental convergence rates with zero liquidity risk and zero transaction costs}
    \label{fig:log-diff}
\end{figure}
The logarithmic difference between the numerical results produced by our proposed scheme and the analytical solution to Merton’s problem is depicted in Figure \ref{fig:log-diff}. This figure illustrates the exponential or even faster convergence rate of our numerical scheme, indicating that accurate numerical results can be achieved efficiently with a limited number of iterations.

\section{Examples}
In this section, three examples featuring different forms of utility functions are presented to illustrate the impact of exogenous and endogenous transaction costs on the investment decisions of investors with varying risk preferences. Note that all calculations are performed using  PyTorch 2.2 on Ubuntu 24.04 LTS, utilizing an NVIDIA RTX 4090 GPU, with a computation time of approximately 20 seconds. It is worth noting that, in most cases, the number of iterations \( k \) required is less than 5 when the relative difference between \( Q_{\phi^{(k)}} \) and \( Q_{\phi^{(k-1)}} \) is sufficiently small, specifically when it falls below \( 10^{-5} \). Unless otherwise mentioned, all of the calculations are carried out for the following parameters:
\begin{table}[H]
    \centering
    \begin{tabular}{cc|cc}
    \toprule
    parameter & value & parameter & value \\
    \midrule
    $r$ & 0.02 & $\bar\theta$ & 0.6 \\
    $\mu$ & 0.05 & $\lambda$ & 5.0 \\
    $\rho_1$ & 0.2 & $\alpha$ & 2.0 \\
    $\rho_2$ & 0.5 & $\beta$ & 0.3 \\
    $\rho_3$ & 0.3 & $\kappa$ & 0.4\% \\
    $\sigma_S$ & 0.4 & $\sigma_L$ & 0.2 \\
    $\delta t$ & $\frac{1}{12}$ & $T$& 1\\
    \bottomrule
    \end{tabular}
    \caption{Default Parameters}
    \label{tab:params}
\end{table}

\subsection{Example 1: Power utility}
We first consider a power utility function, which is defined as:
\begin{equation}
\mathscr{U}(x)=\frac{x^\gamma}{\gamma},
\end{equation}
where $\gamma <1$ and $\gamma \neq 0$ indicates risk aversion. Such a form of utility belongs to the class of constant relative risk aversion (CRRA), as the Arrow-Pratt measure of relative risk aversion is constant:
\begin{equation*}
R(x)=-\frac{\mathscr{U}''(x)}{\mathscr{U}'(x)}x=1-\gamma.
\end{equation*}
This Arrow-Pratt measure implies that the investor is more risk-averse when the value of $\gamma$ is smaller. It should be pointed out that under the power utility function, the stochastic control problem (\ref{Q}) can be simplified to a one-dimensional HJB equation through a change of variables, which facilitates the calculations.

Let $Q(W, L, t)=\frac{W^\gamma}{\gamma}\cdot P(L, t)$, and substitute it into Eq. (\ref{Q}), we have
 \begin{align}
 &\max_{\omega\in [0, 1]} \bigg\{\frac{\partial P}{\partial t}+\bigg(r+(\mu-r)\omega -\sqrt{\frac{2}{\pi\delta t}}\kappa \omega (1-\omega)\sqrt{(\beta L+\sigma_S \rho_1)^2+(1-\rho_1^2)\sigma_S^2}\bigg)\gamma P\\
 &+\frac{\gamma(\gamma-1)}{2}\bigg(\beta^2L^2+\sigma_S^2+2\rho_1\sigma_S\beta L\bigg)\omega^2 P+\bigg(\alpha(\theta-L)+\big(\rho_2\sigma_S+\rho_3\beta L\big)\sigma_L\omega\gamma\bigg)\frac{\partial P}{\partial L}+\frac{\sigma_L^2}{2}\frac{\partial ^2 P}{\partial L^2}\bigg\}=0,\nonumber
 \end{align} 
 with the terminal condition $P(L, T)=1$. Then, we take $\gamma=0.5$ as an example to study the trend of the optimal policy in relation to key parameters of exogenous and endogenous transaction costs.

\begin{figure}[H]
    \centering
    \begin{subfigure}[b]{0.45\textwidth}
        \centering
        \includegraphics[width=\textwidth]{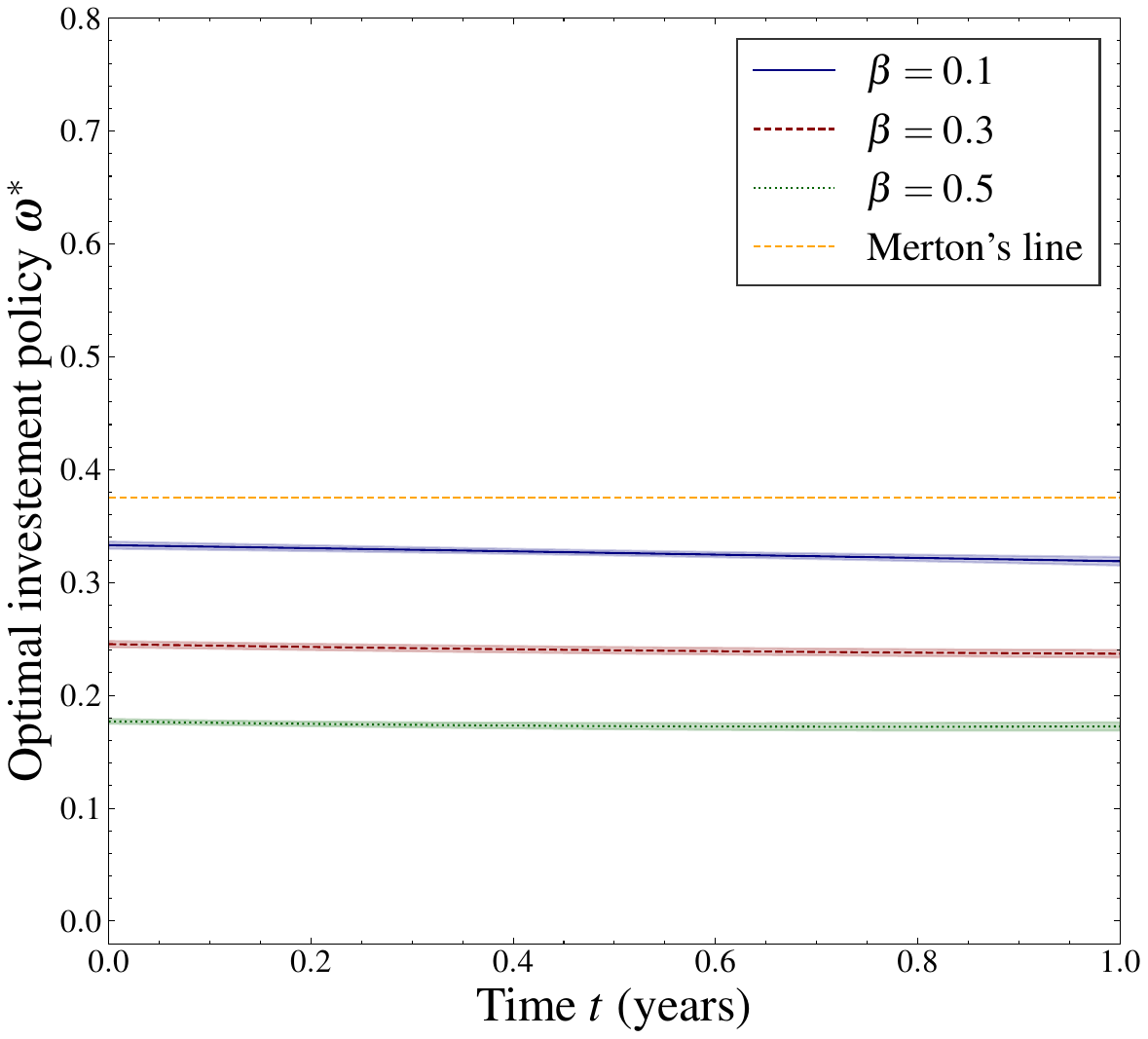}
        \caption{Different $\beta$}
        \label{fig:power-t-beta}
    \end{subfigure}
    \hspace{2em}
    \begin{subfigure}[b]{0.45\textwidth}
        \centering
        \includegraphics[width=\textwidth]{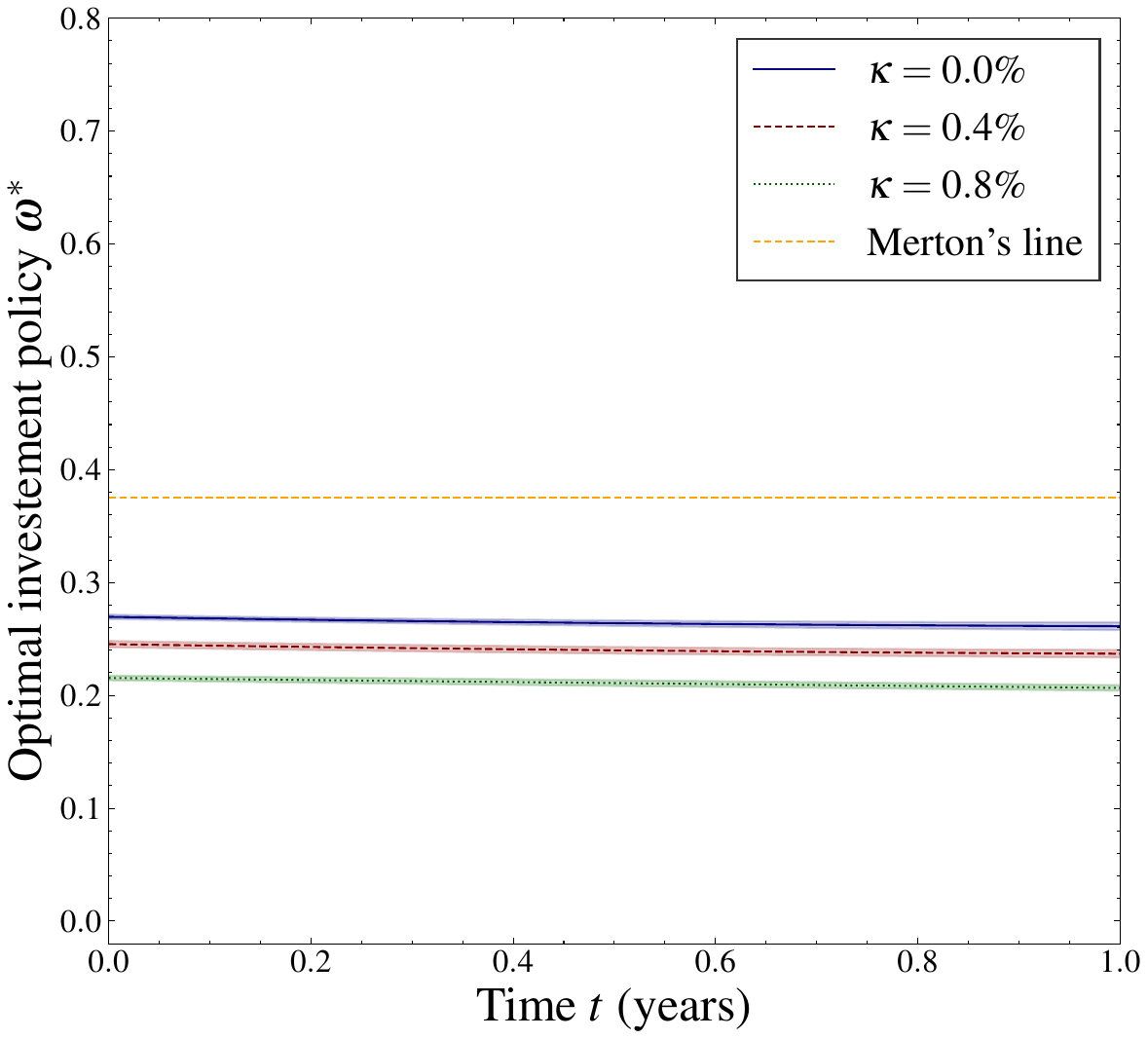}
        \caption{Different $\kappa$}
        \label{fig:power-t-kappa}
    \end{subfigure}
    \caption{The variation with time for $W=2.5$ and $L=0.6$.}
    \label{fig:power-t}
\end{figure}

\begin{figure}[H]
    \centering
    \begin{subfigure}[b]{0.45\textwidth}
        \centering
        \includegraphics[width=\textwidth]{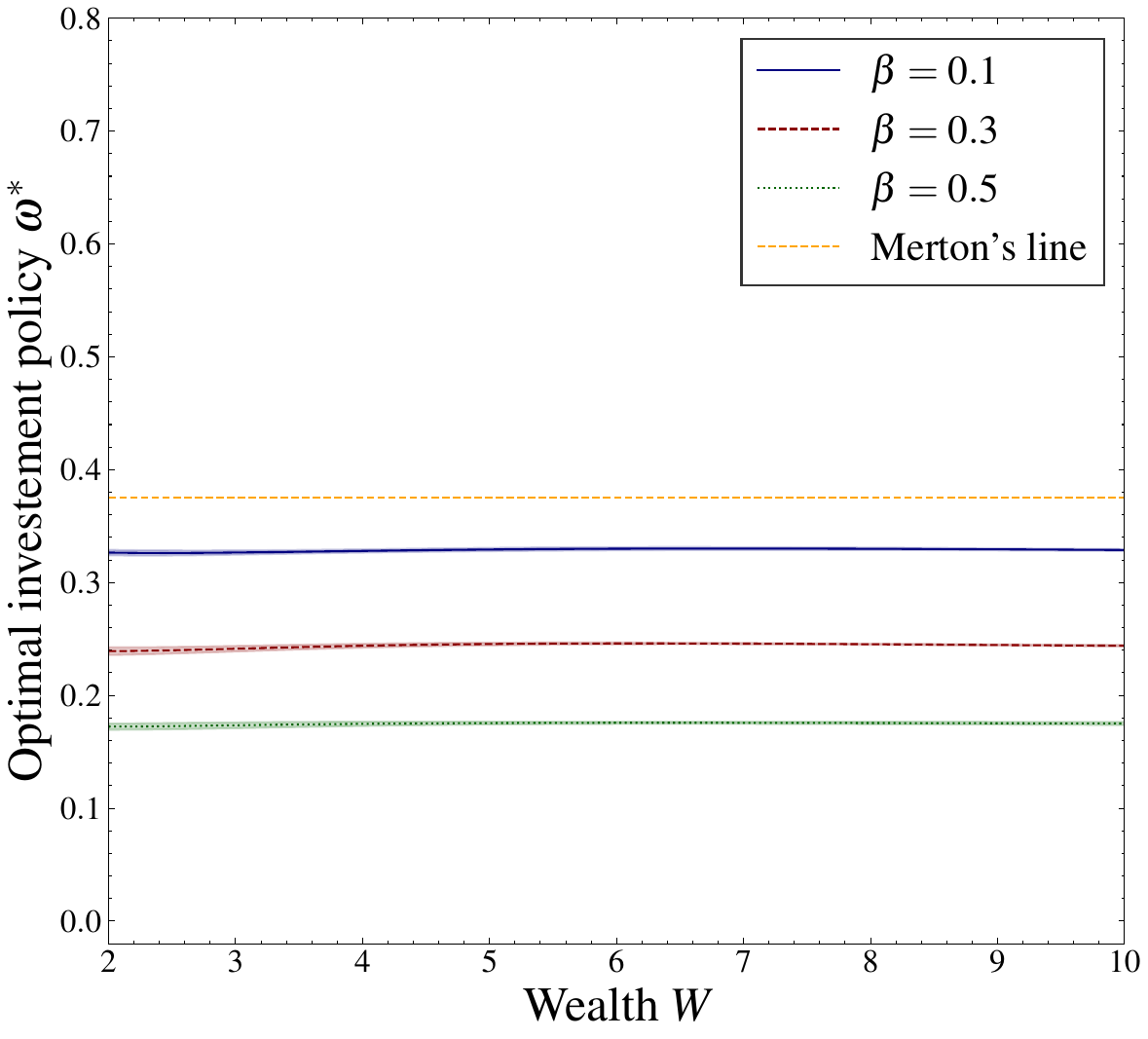}
        \caption{Different $\beta$}
        \label{fig:power-W-beta}
    \end{subfigure}
    \hspace{2em}
    \begin{subfigure}[b]{0.45\textwidth}
        \centering
        \includegraphics[width=\textwidth]{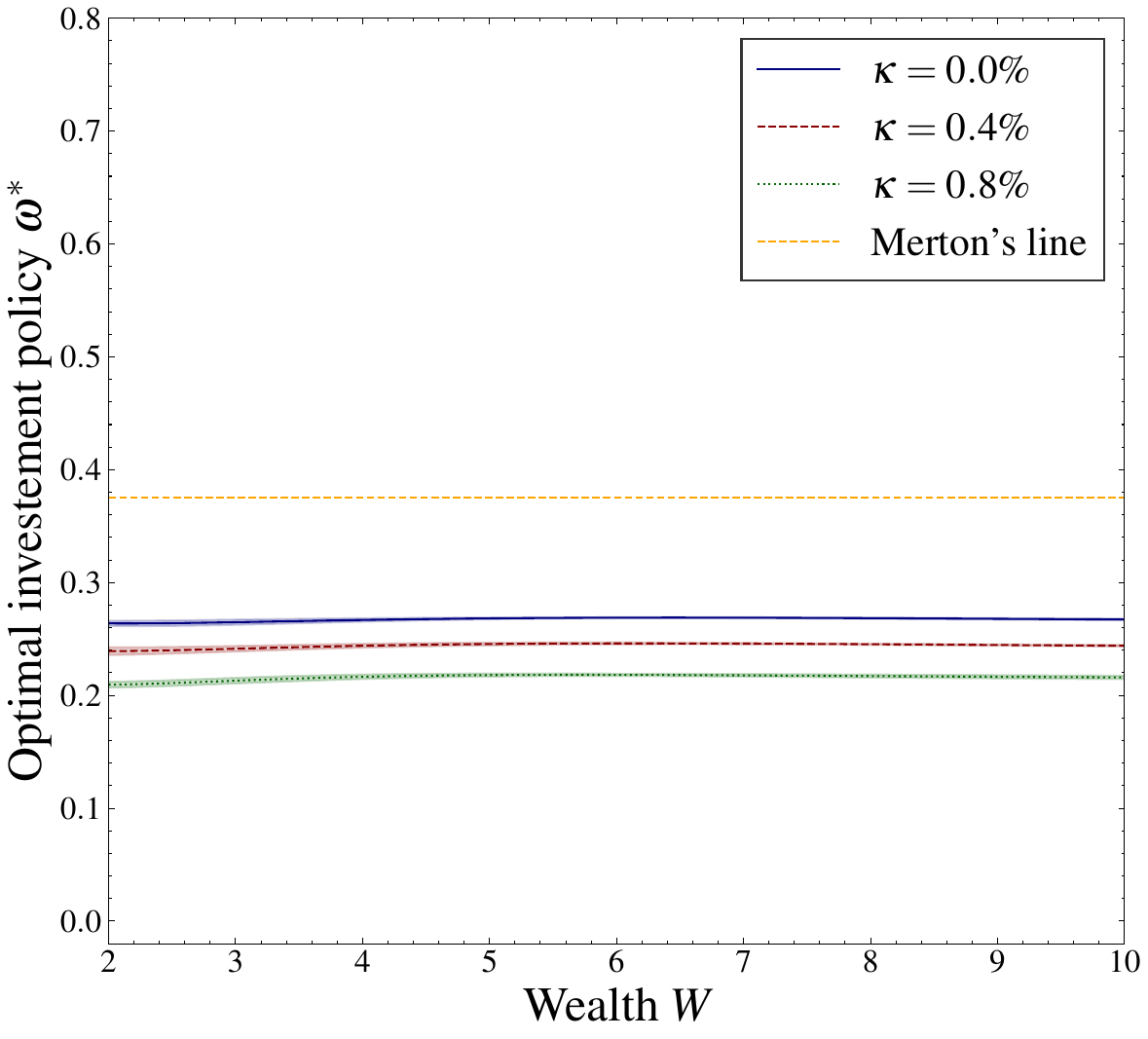}
        \caption{Different $\kappa$}
        \label{fig:power-W-kappa}
    \end{subfigure}
    \caption{The variation with wealth for $L=0.6$ and $t=0.5$.}
    \label{fig:power-W}
\end{figure}

\begin{figure}[H]
    \centering
    \begin{subfigure}[b]{0.45\textwidth}
        \centering
        \includegraphics[width=\textwidth]{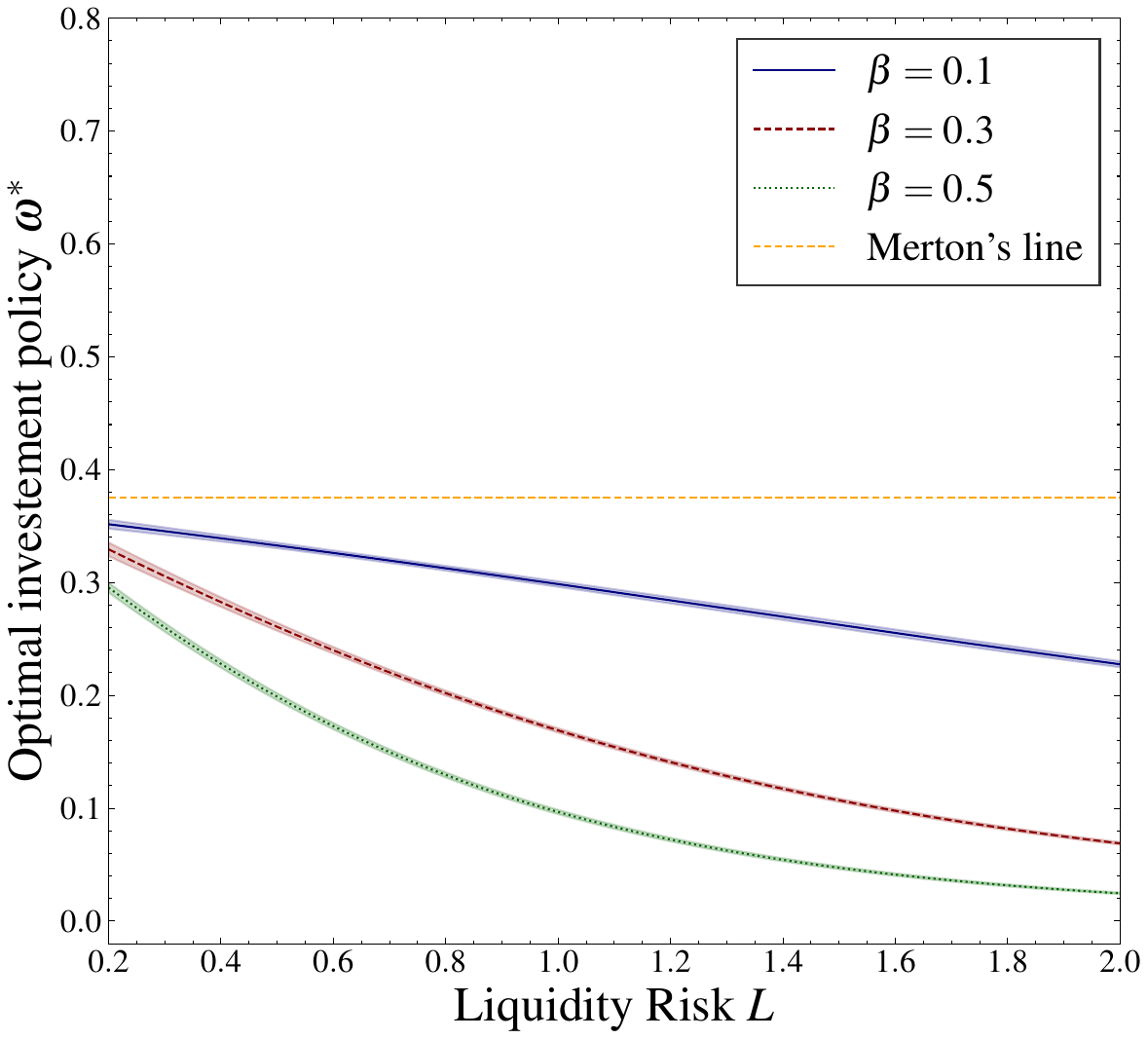}
        \caption{Different $\beta$}
        \label{fig:power-L-beta}
    \end{subfigure}
    \hspace{2em}
    \begin{subfigure}[b]{0.45\textwidth}
        \centering
        \includegraphics[width=\textwidth]{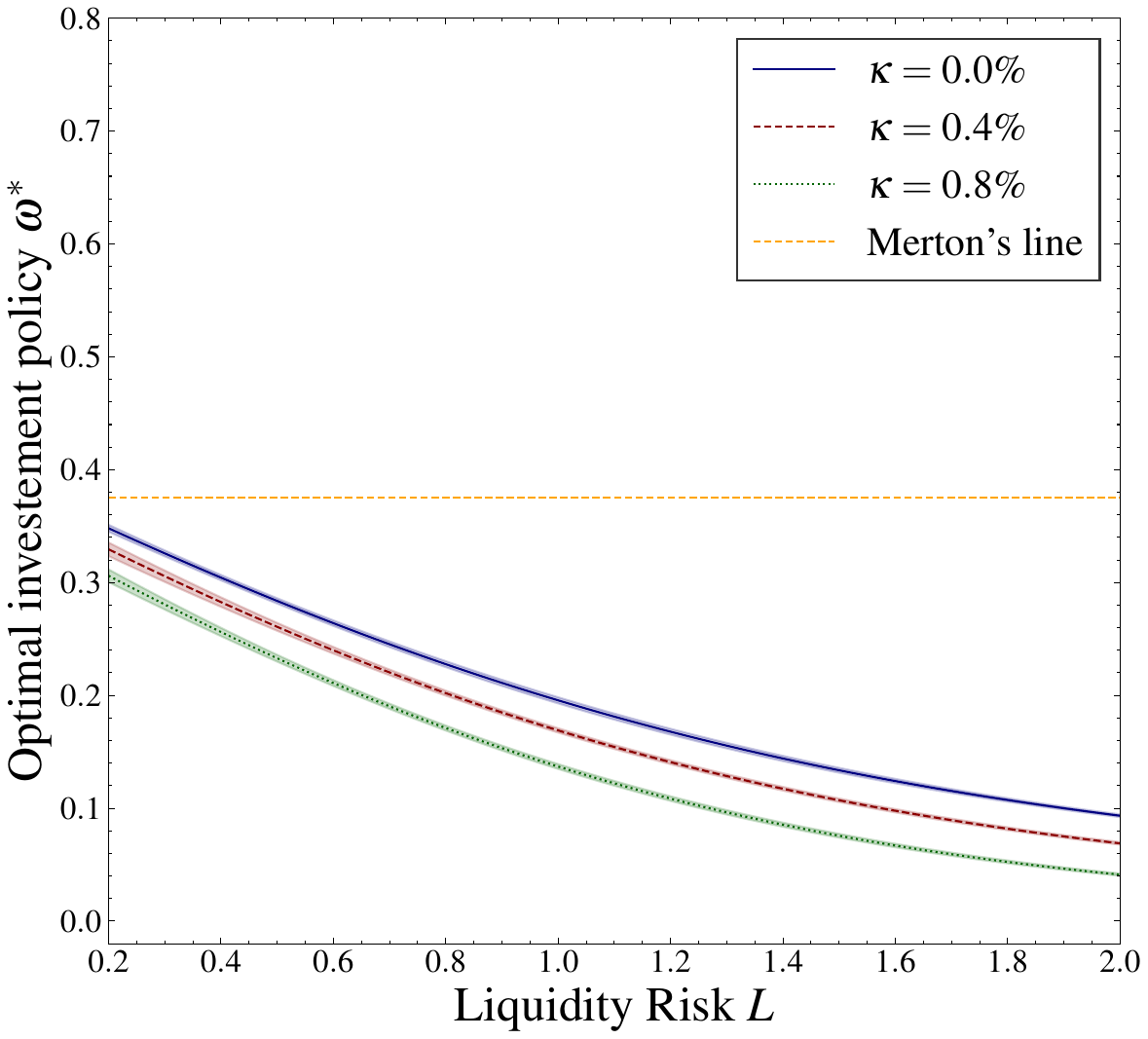}
        \caption{Different $\kappa$}
        \label{fig:power-L-kappa}
    \end{subfigure}
    \caption{The variation with liquidity risk for $W=2.5$ and $t=0.5$.}
    \label{fig:power-L}
\end{figure}

\begin{figure}[H]
    \centering
    \begin{subfigure}[b]{0.3\textwidth}
        \centering
        \includegraphics[width=\textwidth]{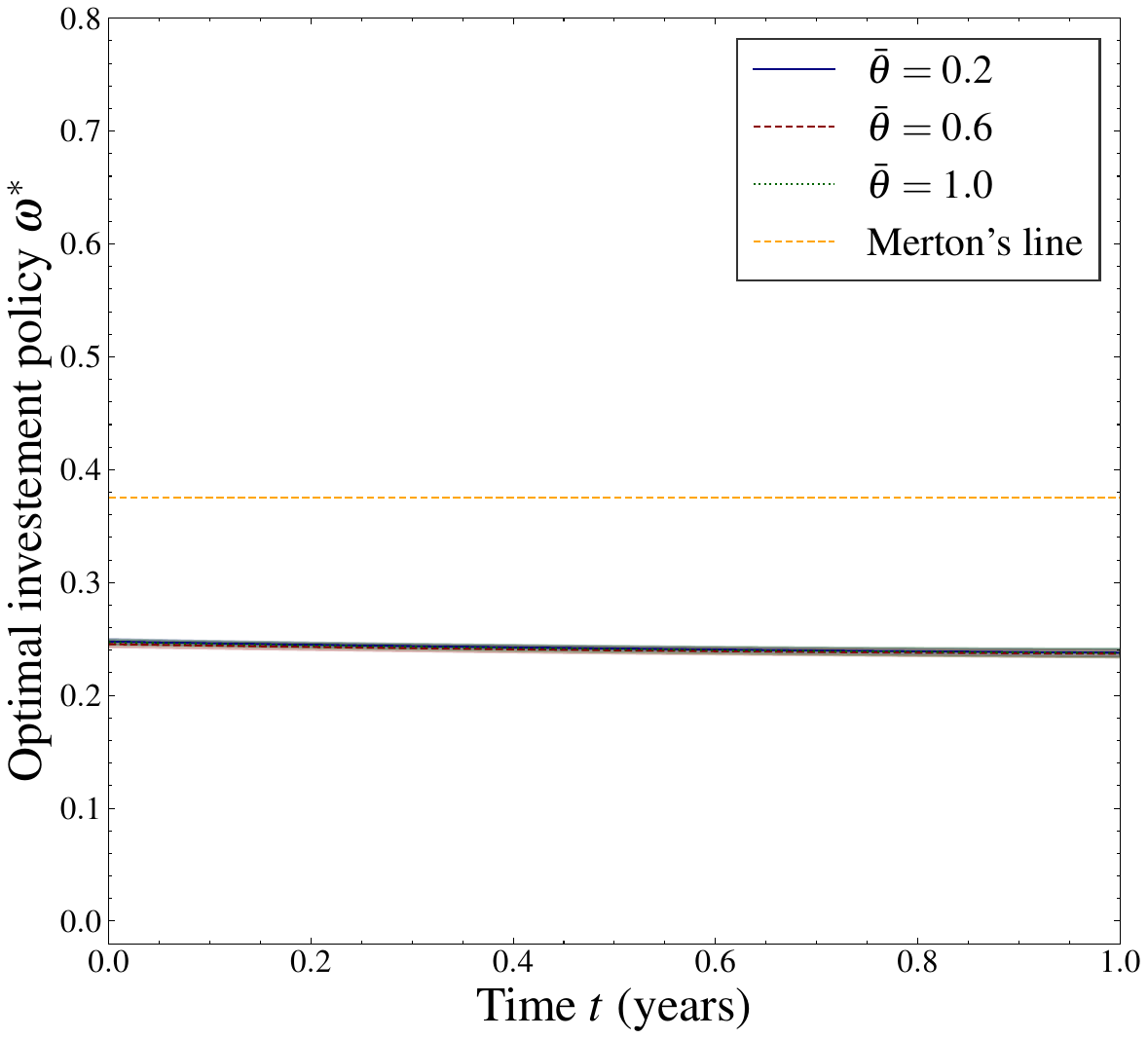}
        \label{fig:power-L-theta-1}
    \end{subfigure}
    \hspace{1em}
    \begin{subfigure}[b]{0.3\textwidth}
        \centering
        \includegraphics[width=\textwidth]{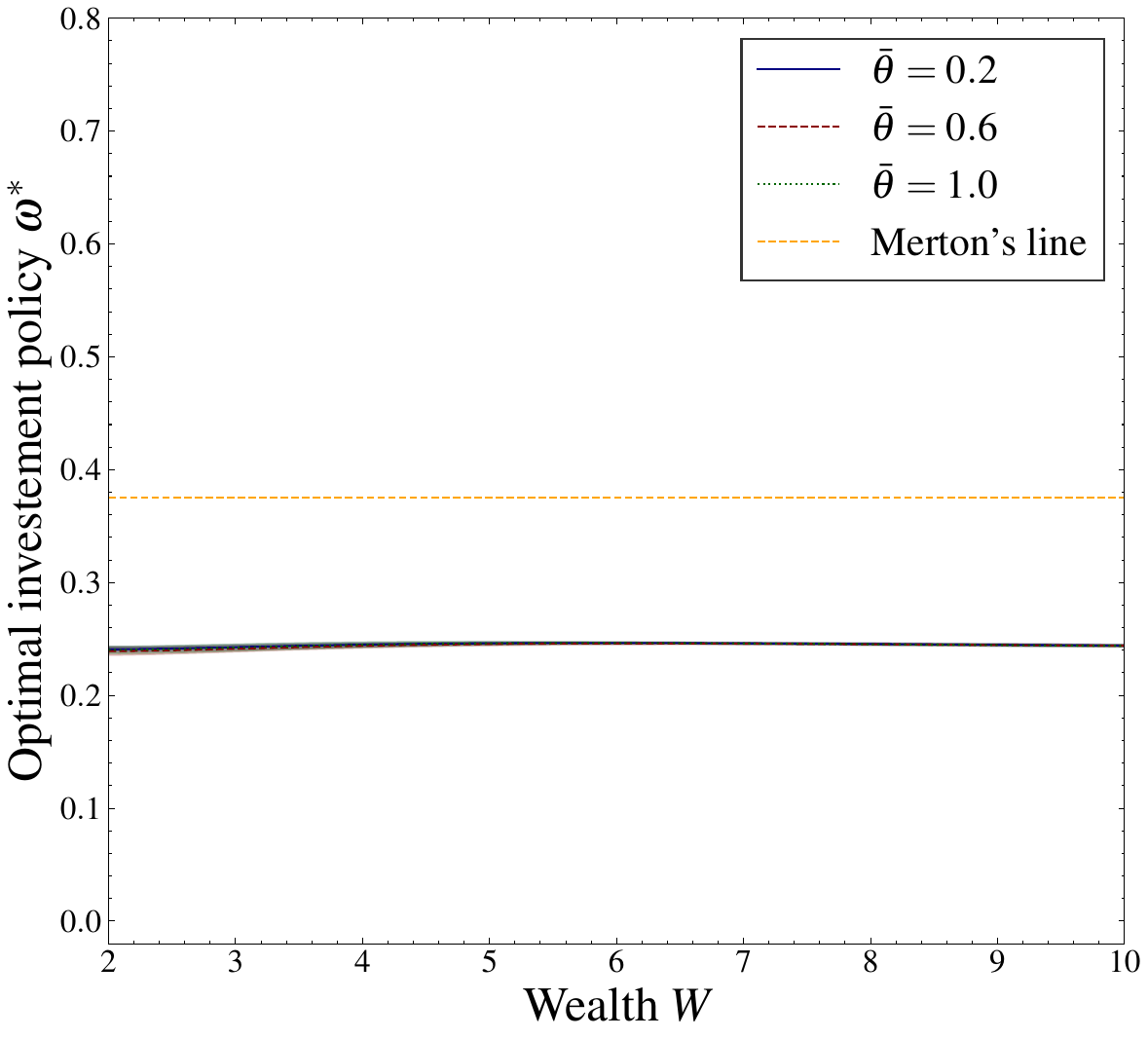}
        \label{fig:power-L-theta-2}
    \end{subfigure}
    \hspace{1em}
    \begin{subfigure}[b]{0.3\textwidth}
        \centering
        \includegraphics[width=\textwidth]{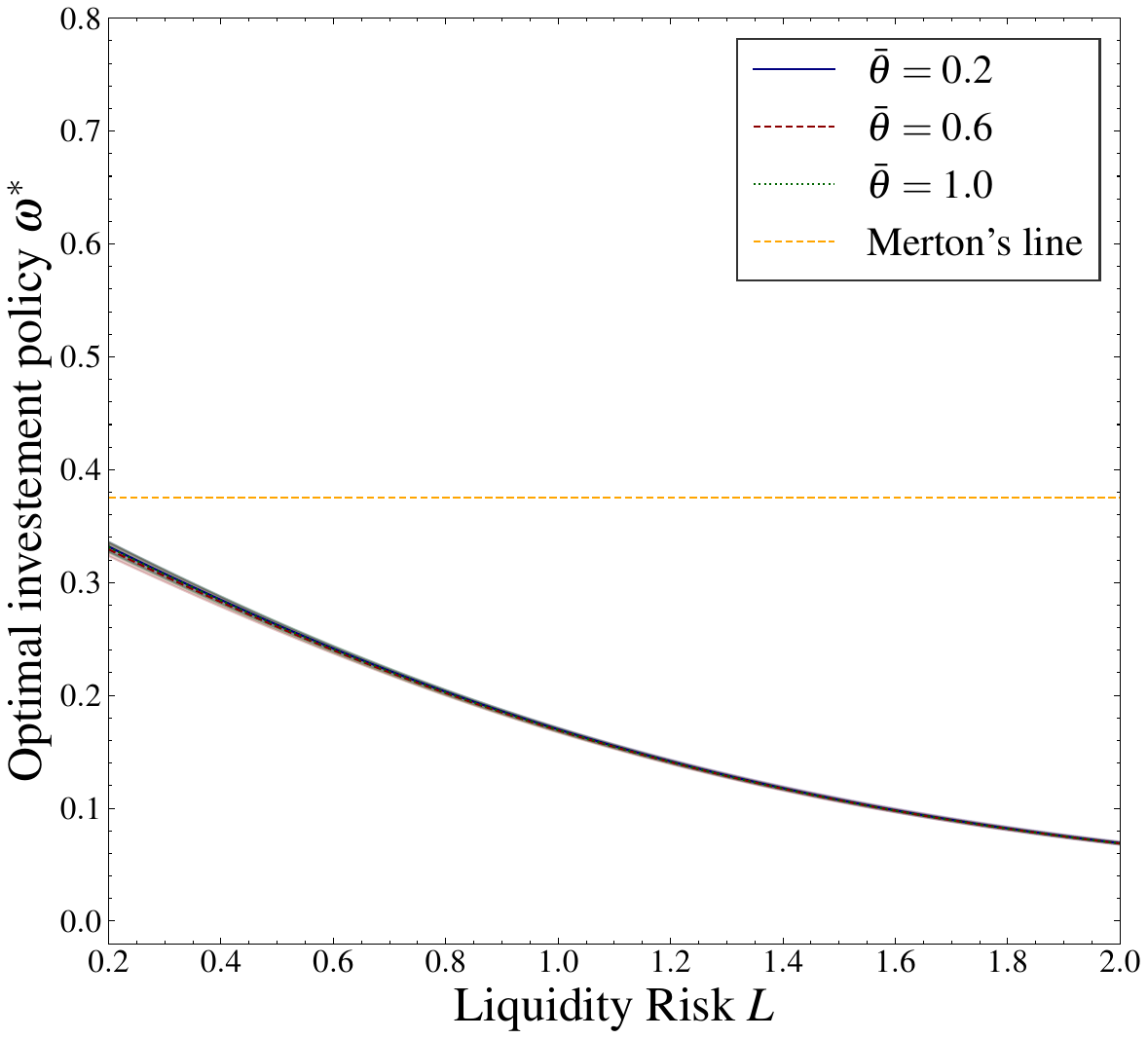}
        \label{fig:power-L-theta-3}
    \end{subfigure}
    \caption{The variation with $\bar{\theta}$ for $W=2.5$, $L=0.6$ and $t=0.5$.}
    \label{fig:power-L-theta}
\end{figure}
Note that the shaded region shown in the figures above represents the results across 10 independent runs, demonstrating the stability and robustness of our algorithm. When either exogenous or endogenous transaction costs are taken into consideration, it affects an investor's portfolio choice, as an investor would invest less in risky assets to incur lower transaction costs. Specifically, when the sensitivity to the level of market liquidity of the asset price or the transaction costs rate increases, the optimal investment policy decreases, and all of these results are below those obtained in Merton's problem without any frictions. As illustrated in Figures \ref{fig:power-t}-\ref{fig:power-L}, it is evident that the optimal investment strategy for risky assets is independent of both time and wealth when such a CRRA utility is considered. However, the level of market liquidity plays a crucial role in portfolio selection. This implies that, when liquidity risk goes higher, investors will carefully and immediately adjust the proportion invested in risky assets. In addition, the results shown in Figure \ref{fig:power-L-theta} demonstrate that the mean-reversion level of liquidity risk, excluding the effects of transaction costs, does not influence an investor's investment decisions. This finding is not surprising, as Kraft \cite{Kraft2005} drawed a similar conclusion regarding the impact of the mean-reversion level of stochastic volatility on portfolio selection under the Heston model \cite{Heston93}.

\subsection{Example 2: Logarithmic utility}
Another classical utility function that belongs to the CRRA class is the logarithmic utility, which is defined as:
\begin{equation}
\mathscr{U}(x)= \ln{x}, 
\end{equation}
and its Arrow–Pratt measure of relative risk aversion is
\begin{equation*}
R(x)=-\frac{\mathscr{U}''(x)}{\mathscr{U}'(x)}x=1.
\end{equation*}
For this case, the investor does not change his attitude towards risk as his wealth varies over time. Additionally, the optimal policy for Merton's problem only depends on an investor's wealth, specifically through the expression $\omega^*= \frac{\mu - r}{\sigma^2}$. Thus, for a fixed value of wealth $W$, the optimal policy for Merton's problem which is denoted as 'Merton's line' remains constant as shown in the figures below.

\begin{figure}[H]
    \centering
    \begin{subfigure}[b]{0.45\textwidth}
        \centering
        \includegraphics[width=\textwidth]{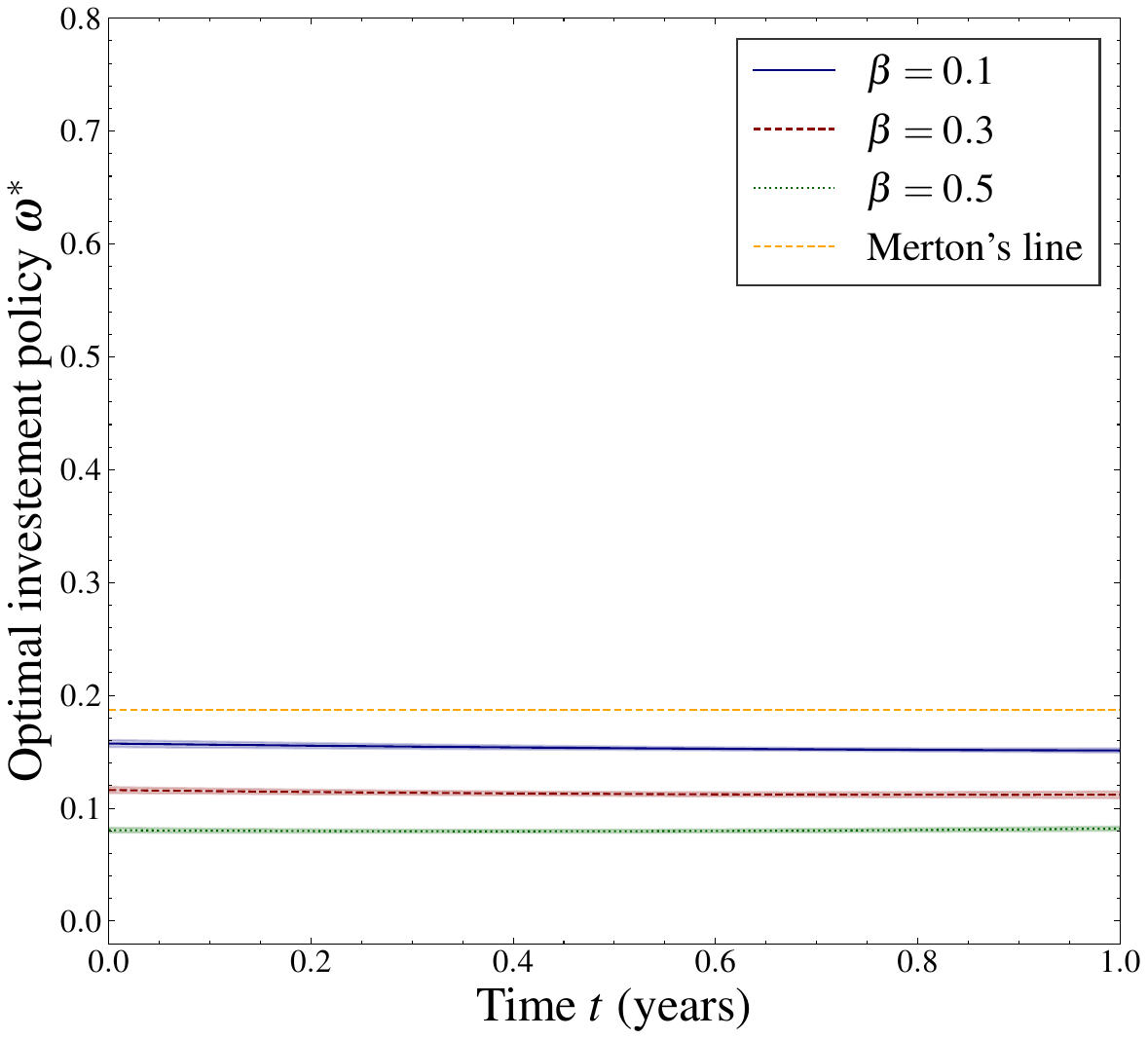}
        \caption{Different $\beta$}
        \label{fig:log-t-beta}
    \end{subfigure}
    \hspace{2em}
    \begin{subfigure}[b]{0.45\textwidth}
        \centering
        \includegraphics[width=\textwidth]{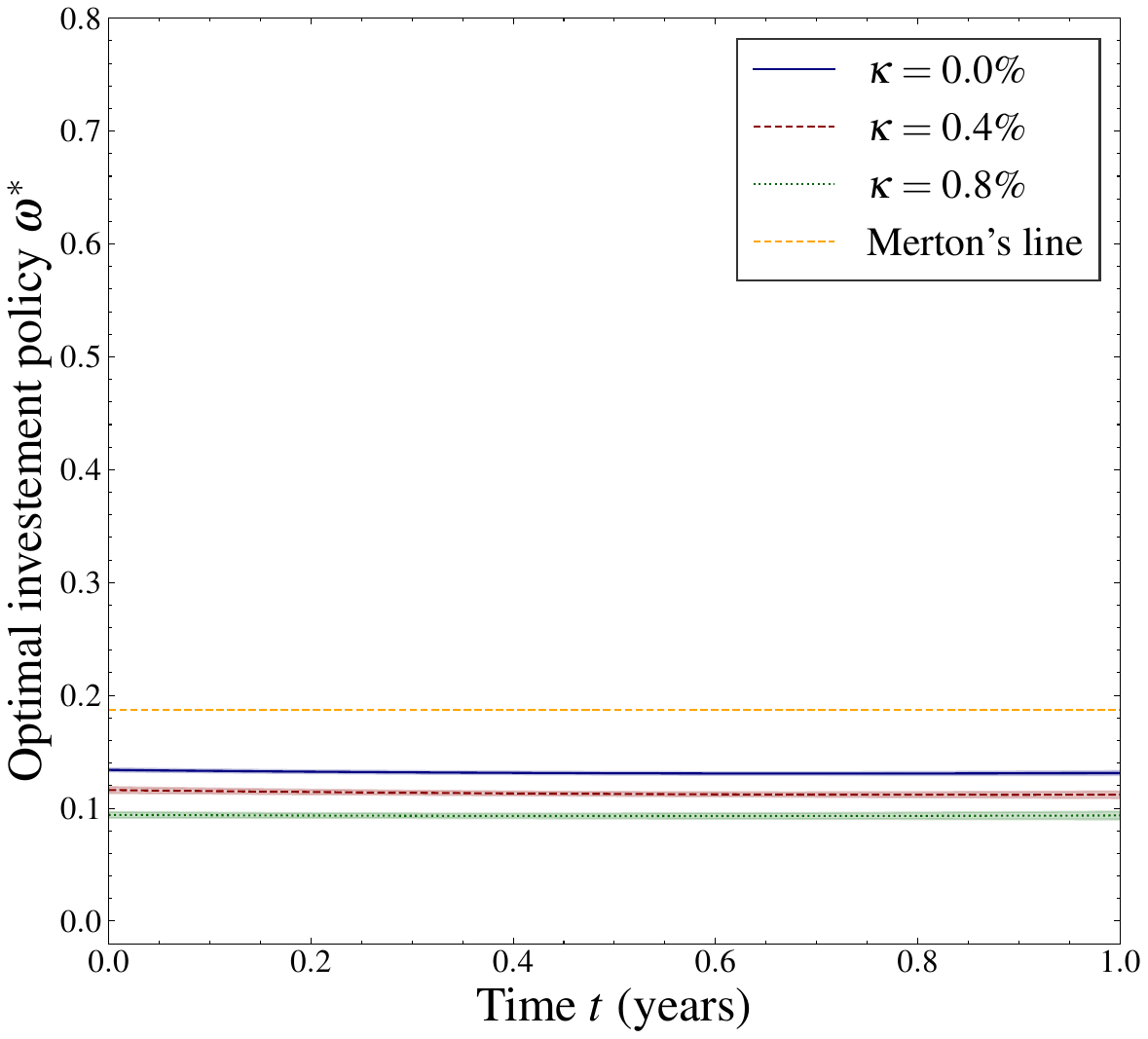}
        \caption{Different $\kappa$}
        \label{fig:log-t-kappa}
    \end{subfigure}
    \caption{The variation with time for $W=2.5$ and $L=0.6$.}
    \label{fig:log-t}
\end{figure}

\begin{figure}[H]
    \centering
    \begin{subfigure}[b]{0.45\textwidth}
        \centering
        \includegraphics[width=\textwidth]{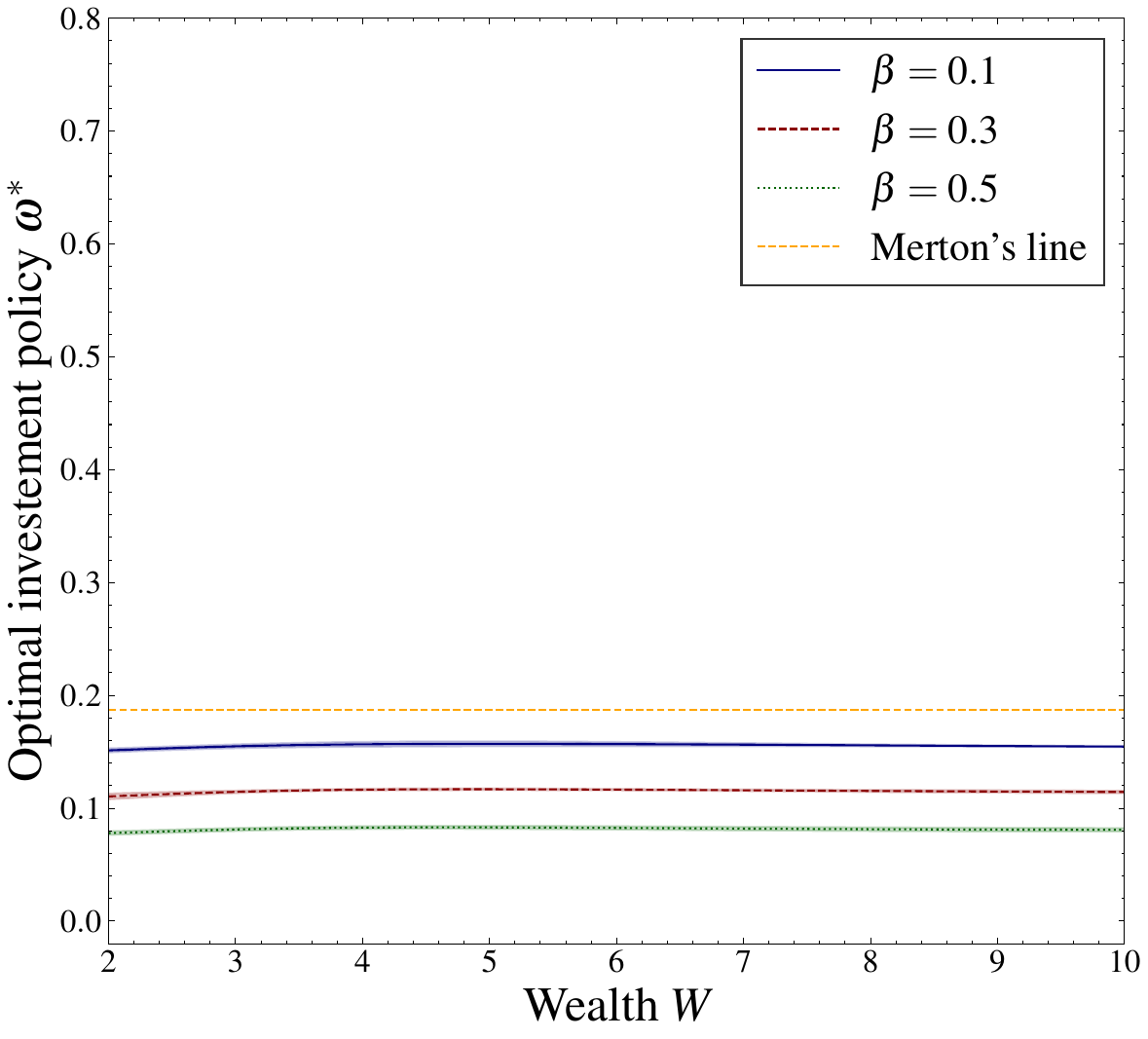}
        \caption{Different $\beta$}
        \label{fig:log-W-beta}
    \end{subfigure}
    \hspace{2em}
    \begin{subfigure}[b]{0.45\textwidth}
        \centering
        \includegraphics[width=\textwidth]{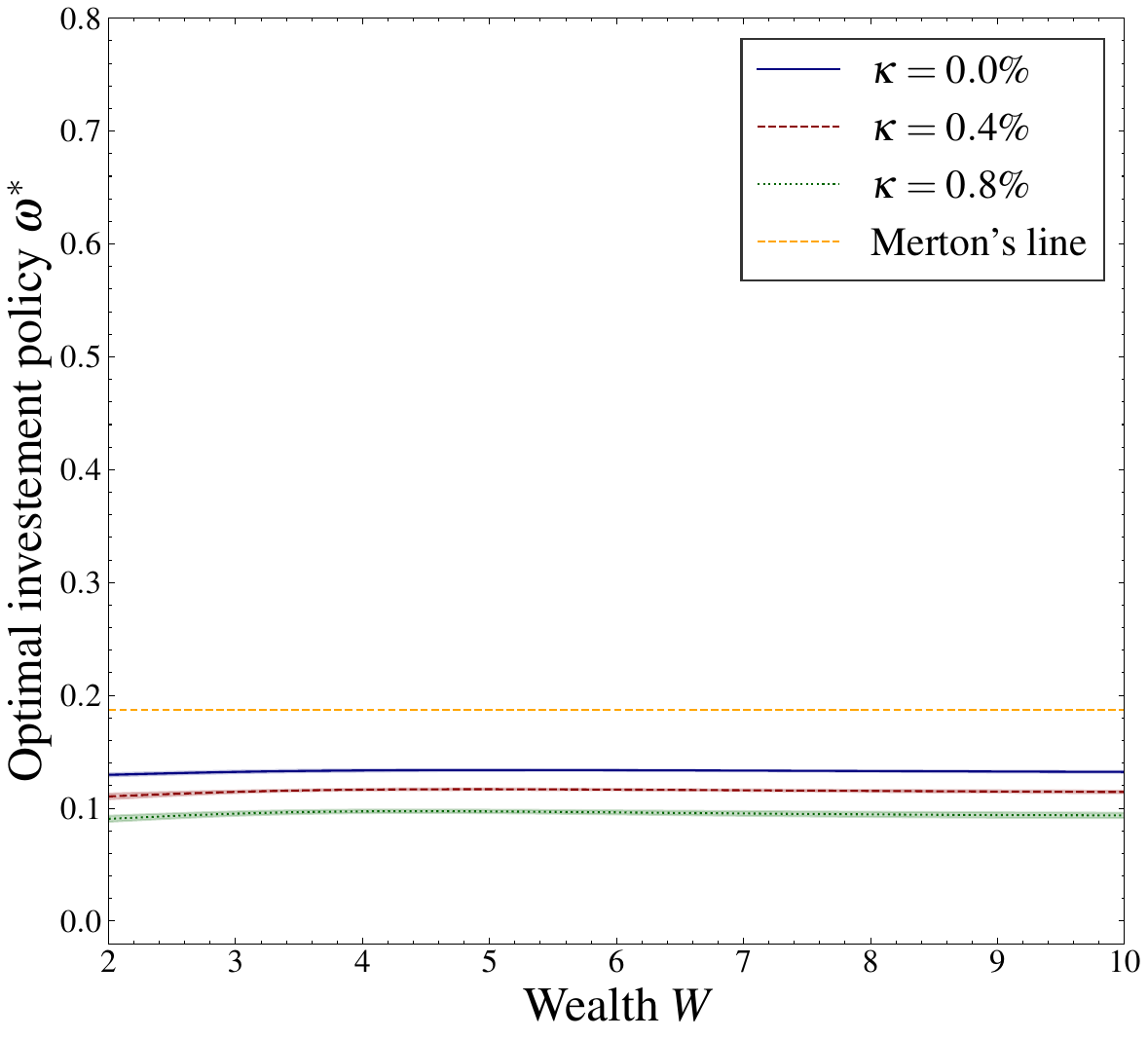}
        \caption{Different $\kappa$}
        \label{fig:log-W-kappa}
    \end{subfigure}
    \caption{The variation with wealth for $L=0.6$ and $t=0.5$.}
    \label{fig:log-W}
\end{figure}

\begin{figure}[H]
    \centering
    \begin{subfigure}[b]{0.45\textwidth}
        \centering
        \includegraphics[width=\textwidth]{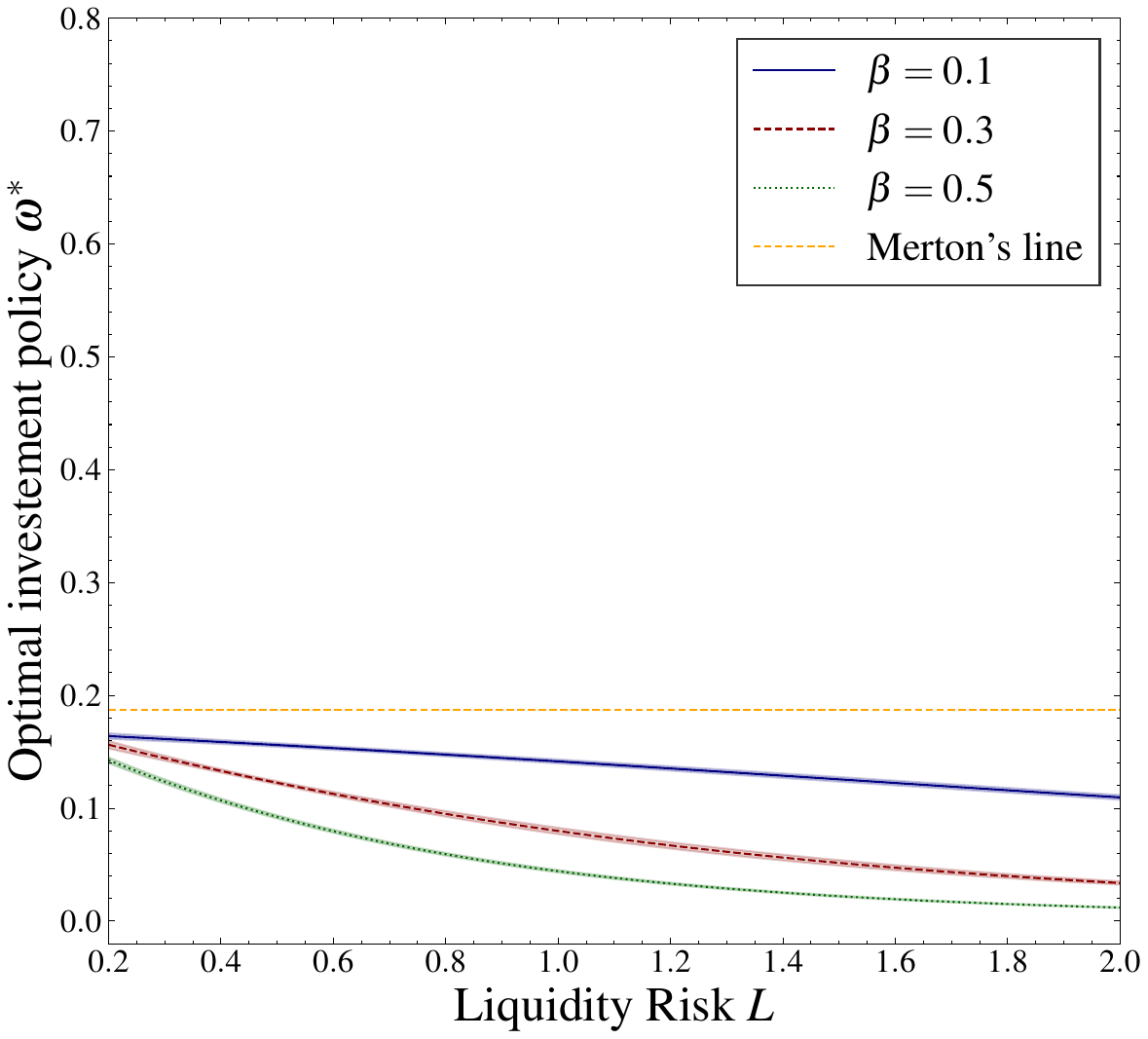}
        \caption{Different $\beta$}
        \label{fig:log-L-beta}
    \end{subfigure}
    \hspace{2em}
    \begin{subfigure}[b]{0.45\textwidth}
        \centering
        \includegraphics[width=\textwidth]{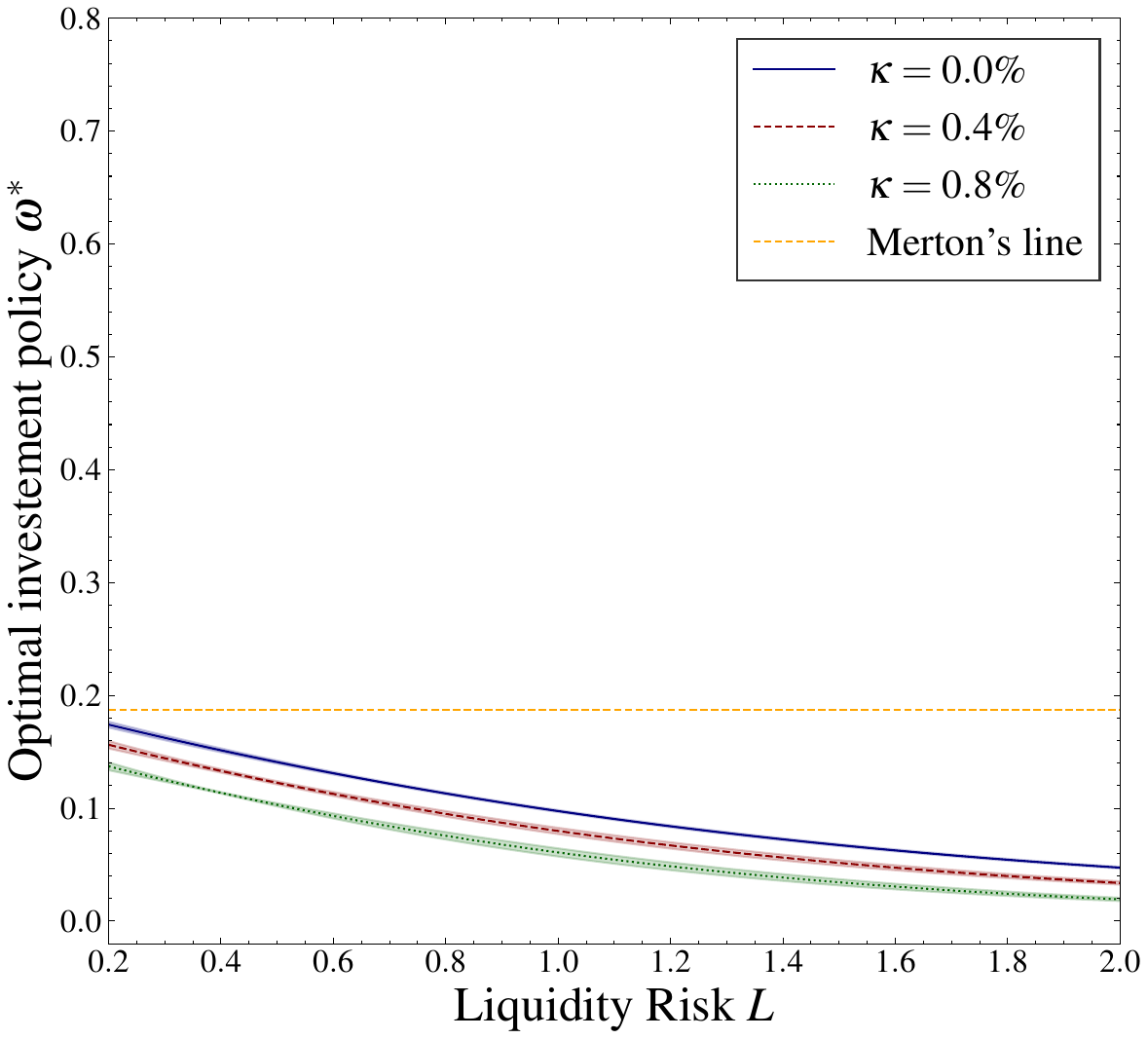}
        \caption{Different $\kappa$}
        \label{fig:log-L-kappa}
    \end{subfigure}
    \caption{The variation with liquidity risk for $W=2.5$ and $t=0.5$.}
    \label{fig:log-L}
\end{figure}

\begin{figure}[H]
    \centering
    \begin{subfigure}[b]{0.3\textwidth}
        \centering
        \includegraphics[width=\textwidth]{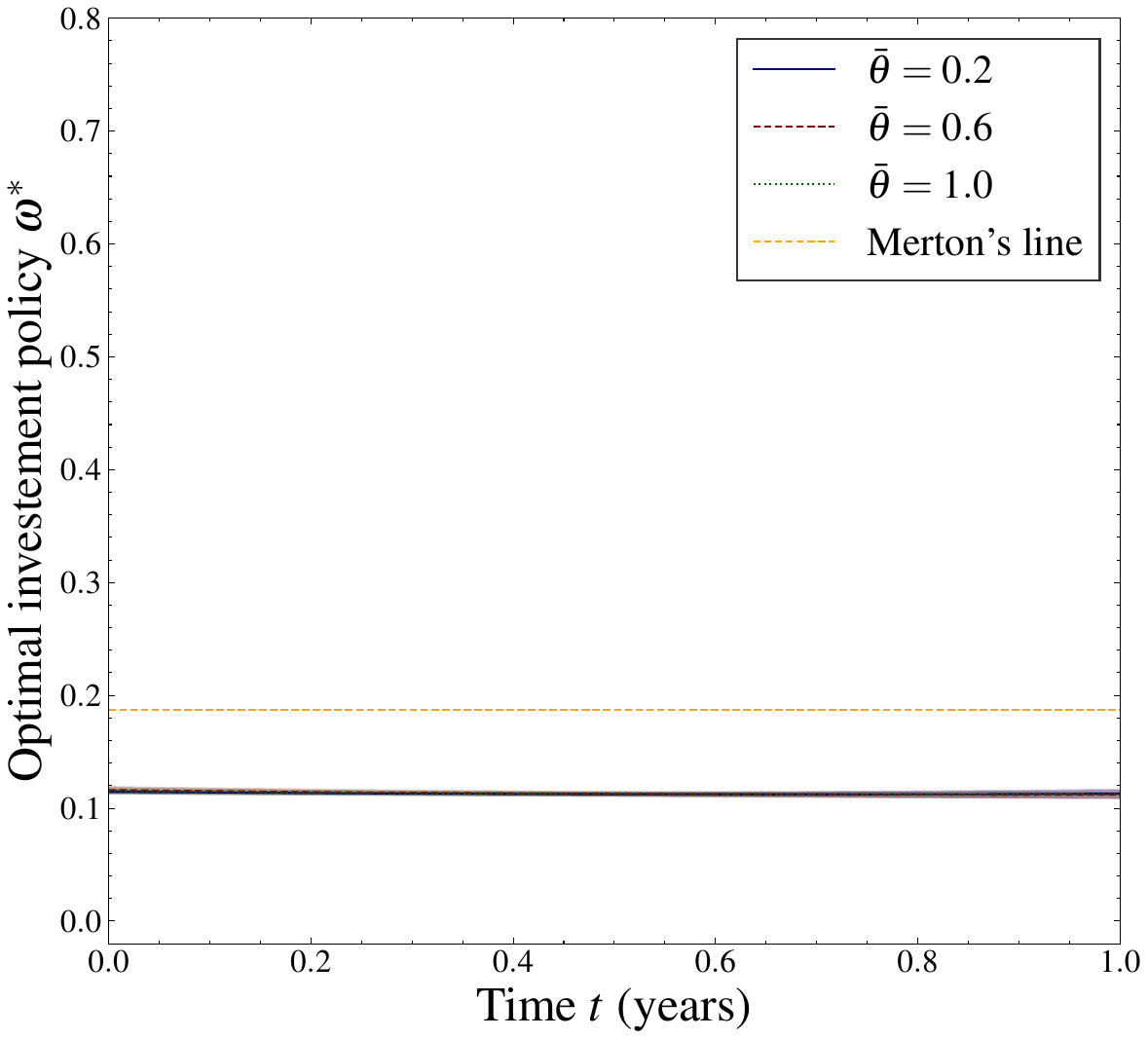}
        \label{fig:log-L-theta-1}
    \end{subfigure}
    \hspace{1em}
    \begin{subfigure}[b]{0.3\textwidth}
        \centering
        \includegraphics[width=\textwidth]{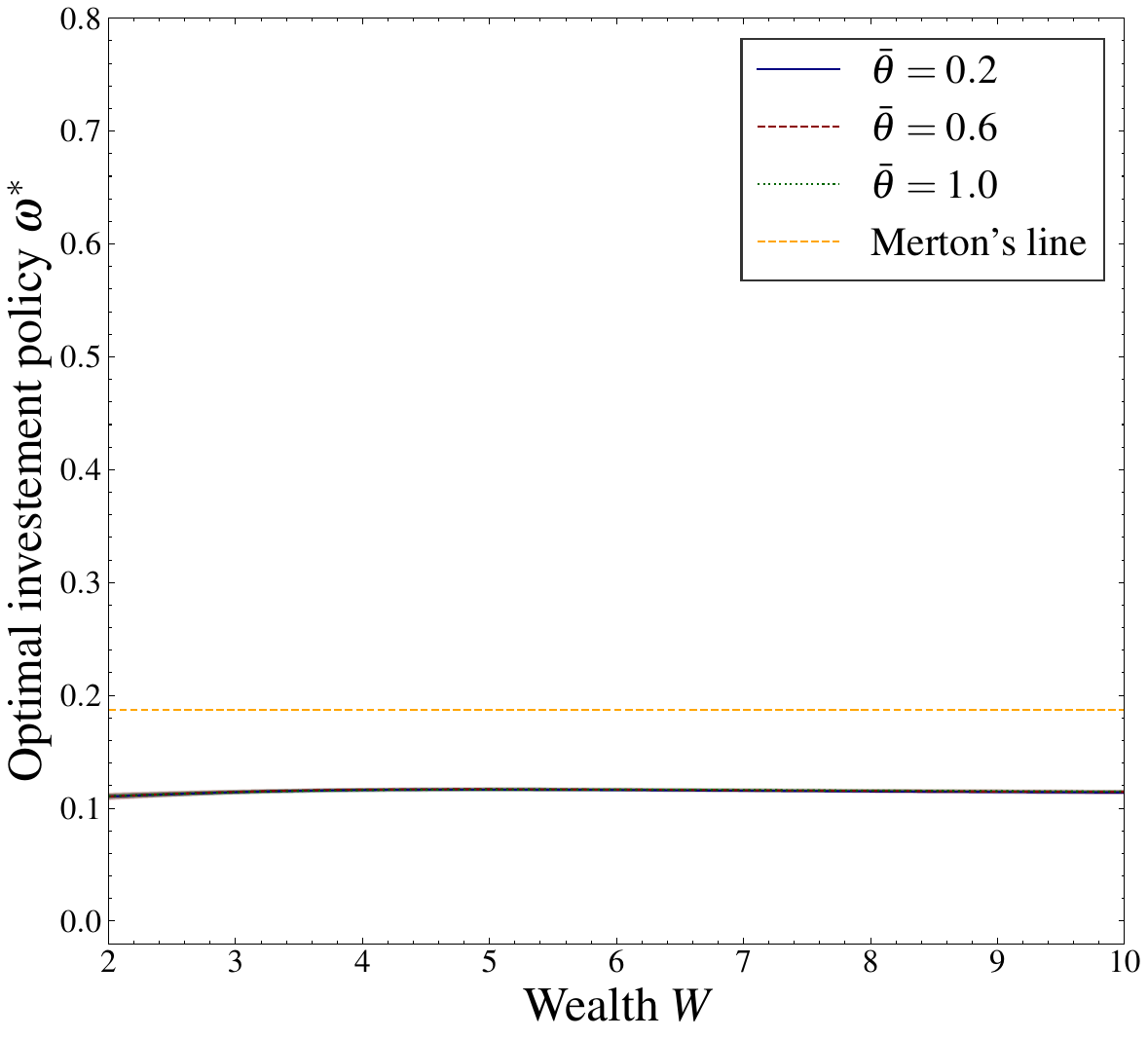}
        \label{fig:log-L-theta-2}
    \end{subfigure}
    \hspace{1em}
    \begin{subfigure}[b]{0.3\textwidth}
        \centering
        \includegraphics[width=\textwidth]{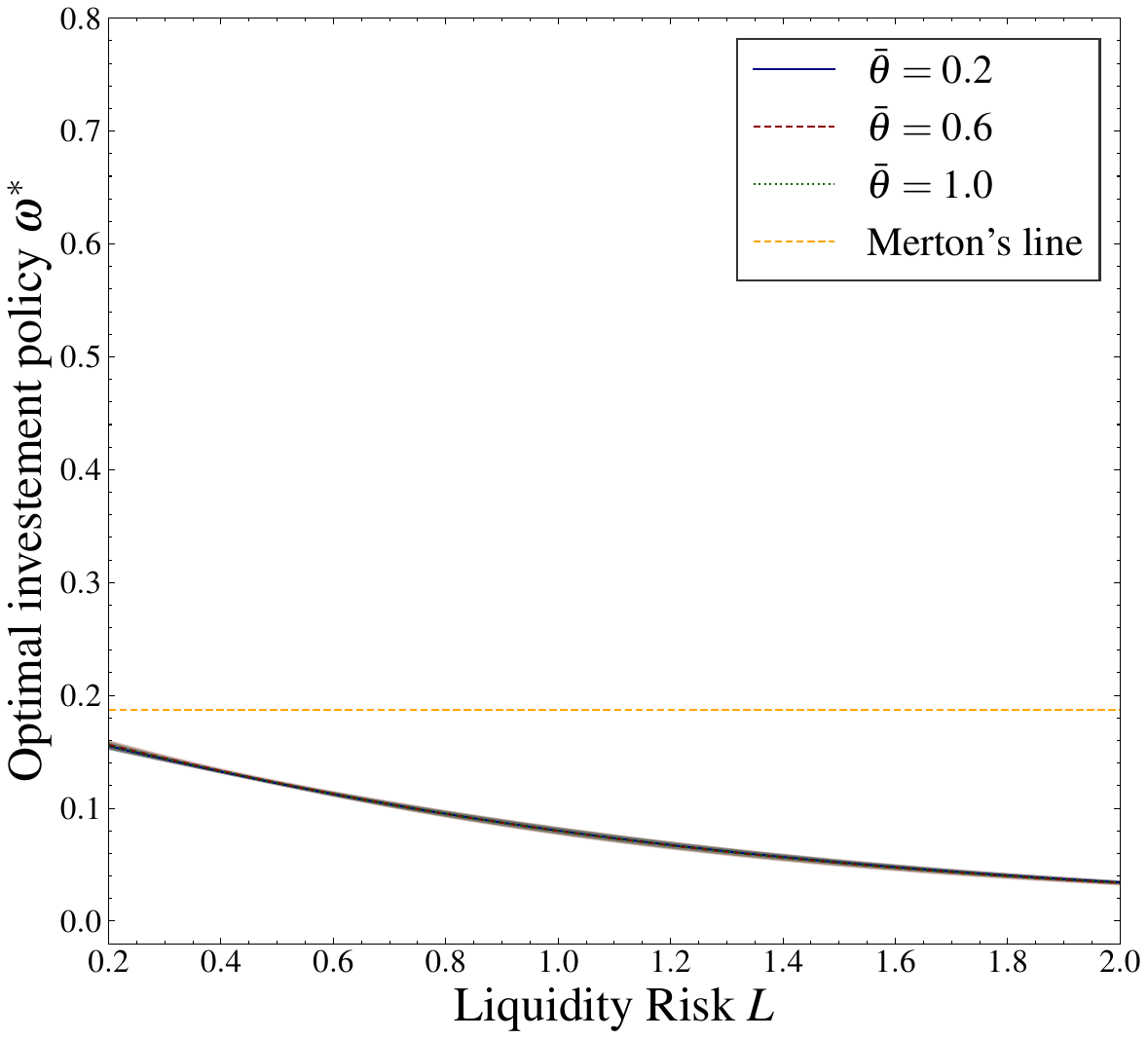}
        \label{fig:log-L-theta-3}
    \end{subfigure}
    \caption{The variation with $\bar{\theta}$ for $W=2.5$, $L=0.6$ and $t=0.5$.}
    \label{fig:log-L-theta}
\end{figure}

In this case, the results are similar to those of Example 1. A larger value of $\beta$, reflecting an increased effect of liquidity risk on the underlying asset, correlates with a reduced proportion of the portfolio invested in risky assets. Similarly, a higher transaction cost rate $\kappa$, which directly increases exogenous costs, leads to a decreased investment in risky assets.

\subsection{Example 3: Exponential utility}
\begin{figure}[H]
    \centering
    \begin{subfigure}[b]{0.45\textwidth}
        \centering
        \includegraphics[width=\textwidth]{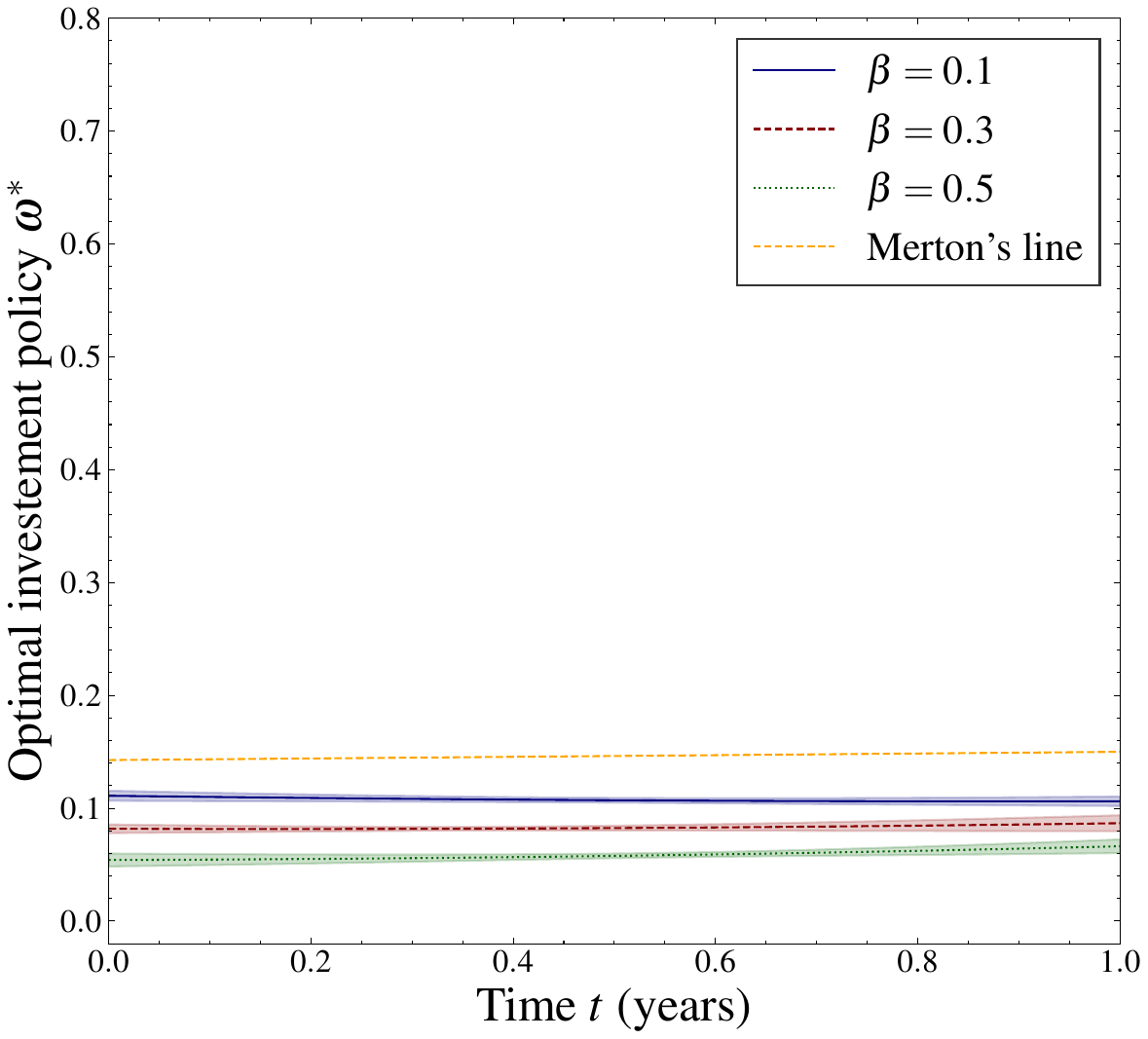}
        \caption{Different $\beta$}
        \label{fig:exp-t-beta}
    \end{subfigure}
    \hspace{2em}
    \begin{subfigure}[b]{0.45\textwidth}
        \centering
        \includegraphics[width=\textwidth]{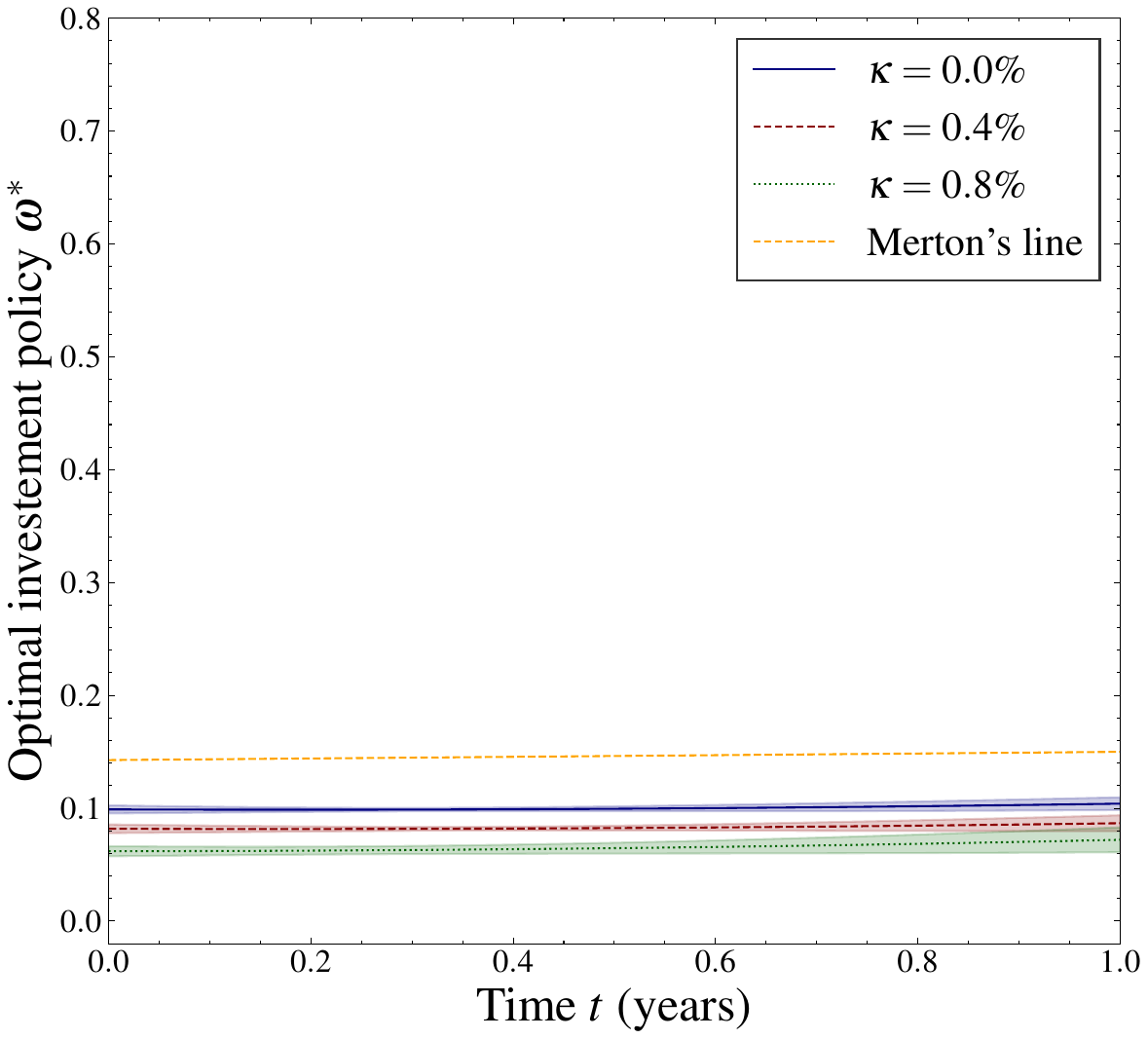}
        \caption{Different $\kappa$}
        \label{fig:exp-t-kappa}
    \end{subfigure}
    \caption{The variation with time for $W=2.5$ and $L=0.6$.}
    \label{fig:expr-t}
\end{figure}

\begin{figure}[H]
    \centering
    \begin{subfigure}[b]{0.45\textwidth}
        \centering
        \includegraphics[width=\textwidth]{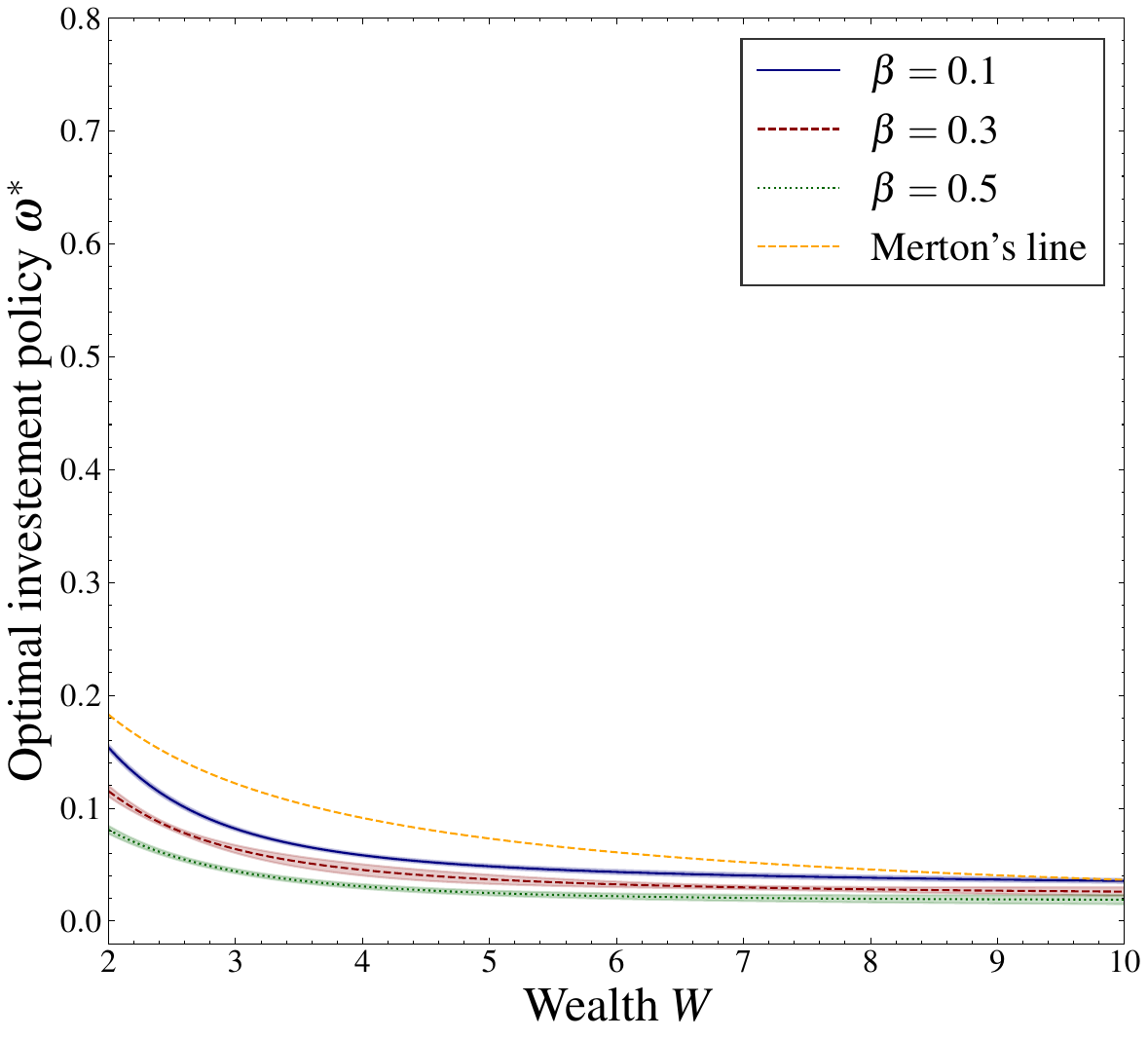}
        \caption{Different $\beta$}
        \label{fig:exp-W-beta}
    \end{subfigure}
    \hspace{2em}
    \begin{subfigure}[b]{0.45\textwidth}
        \centering
        \includegraphics[width=\textwidth]{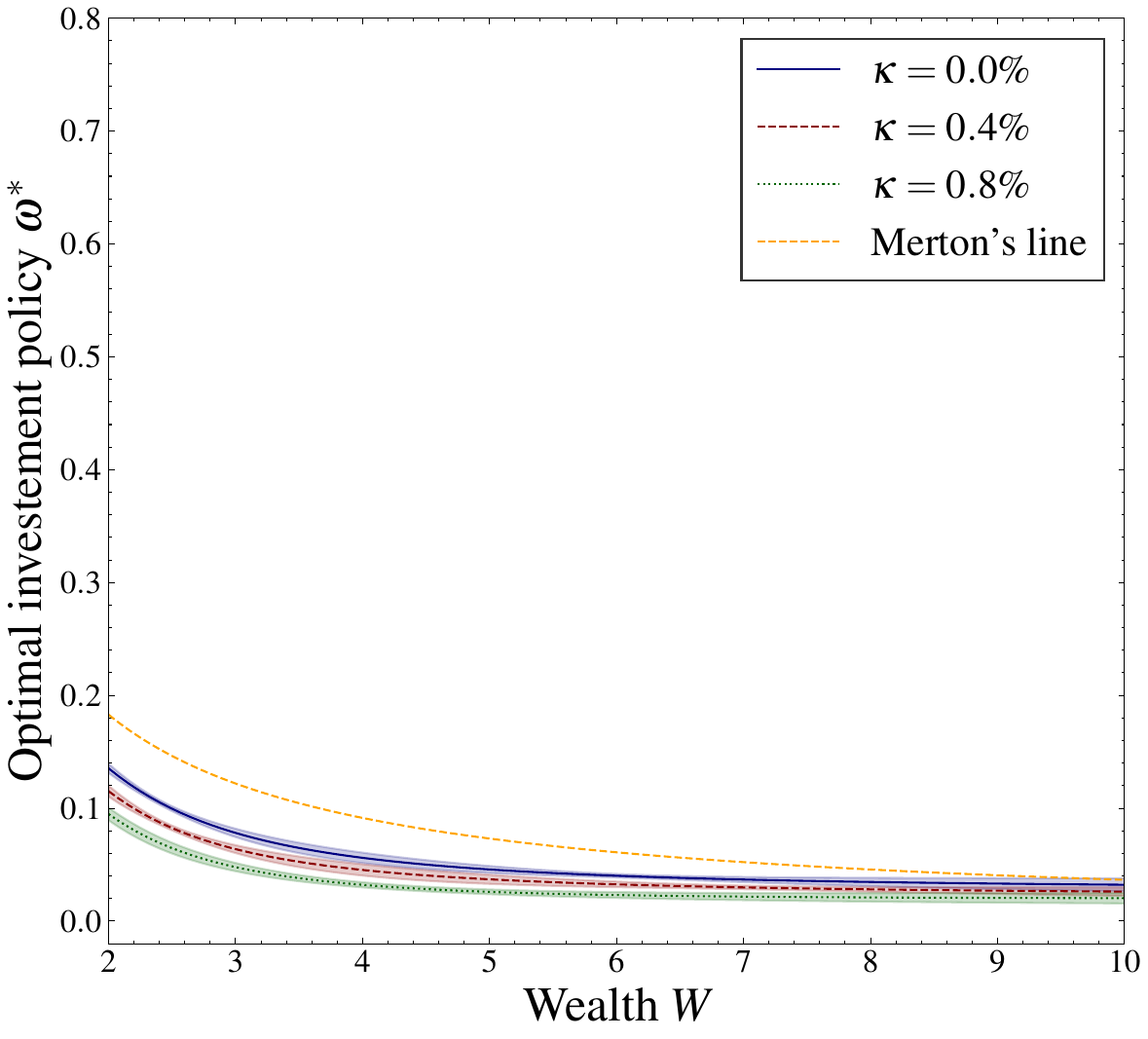}
        \caption{Different $\kappa$}
        \label{fig:exp-W-kappa}
    \end{subfigure}
    \caption{The variation with wealth for $L=0.6$ and $t=0.5$.}
    \label{fig:exp-W}
\end{figure}

In this section, we discuss the case involving non-CRRA utility function to study the impact of transaction costs. Due to its convenience for assessment, the exponential utility function is one of the most commonly used, and a typical form is
\begin{equation}
\mathscr{U}(x)=1-e^{-\eta x},
\end{equation}
where $\eta>0$ determines the curvature of the function and consequently, the degree of risk aversion of the decision maker. The Arrow-Pratt measure of absolute risk aversion for the exponential utility is 
\begin{equation*}
A(x)=-\frac{\mathscr{U}''(x)}{\mathscr{U}'(x)}=\eta,
\end{equation*}
which implies that it belong to the CARA class. And $R(x)=\eta x$ indicates that the investor shows more risk-aversion as his or her wealth becomes larger. For this case, the 'Merton's Line' as presented in Figures \ref{fig:expr-t} and \ref{fig:exp-W}, exhibits an increasing trend with respect to time $t$ and a decreasing trend with respect to an investor's wealth $W$, since $\omega^*= e^{-r(T-t)}\frac{\mu -r}{\eta \sigma^2} \frac{1}{W}$. For the following calculations, we set the level of risk aversion to 0.5.

\begin{figure}[H]
    \centering
    \begin{subfigure}[b]{0.45\textwidth}
        \centering
        \includegraphics[width=\textwidth]{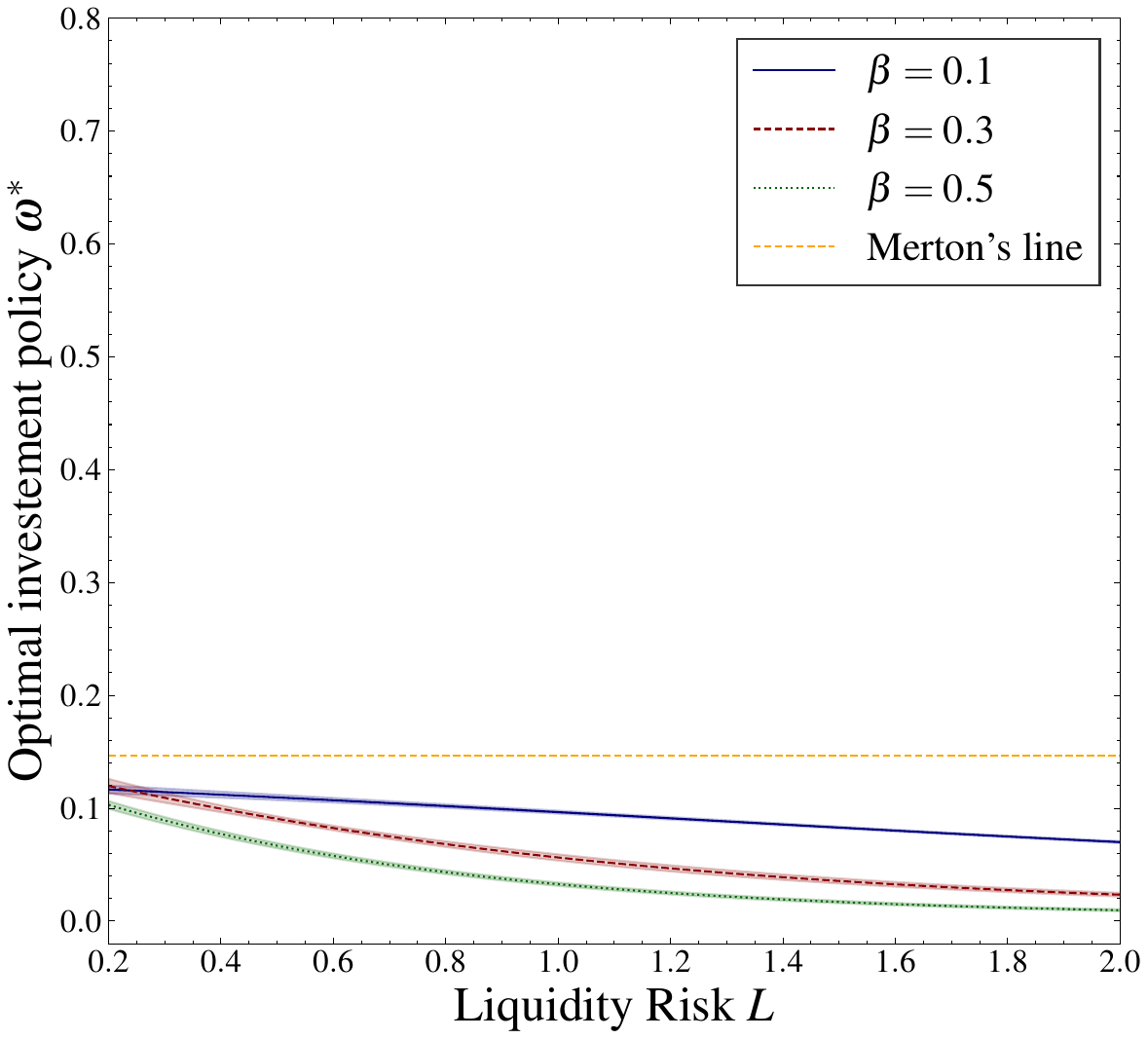}
        \caption{Different $\beta$}
        \label{fig:exp-L-beta}
    \end{subfigure}
    \hspace{2em}
    \begin{subfigure}[b]{0.45\textwidth}
        \centering
        \includegraphics[width=\textwidth]{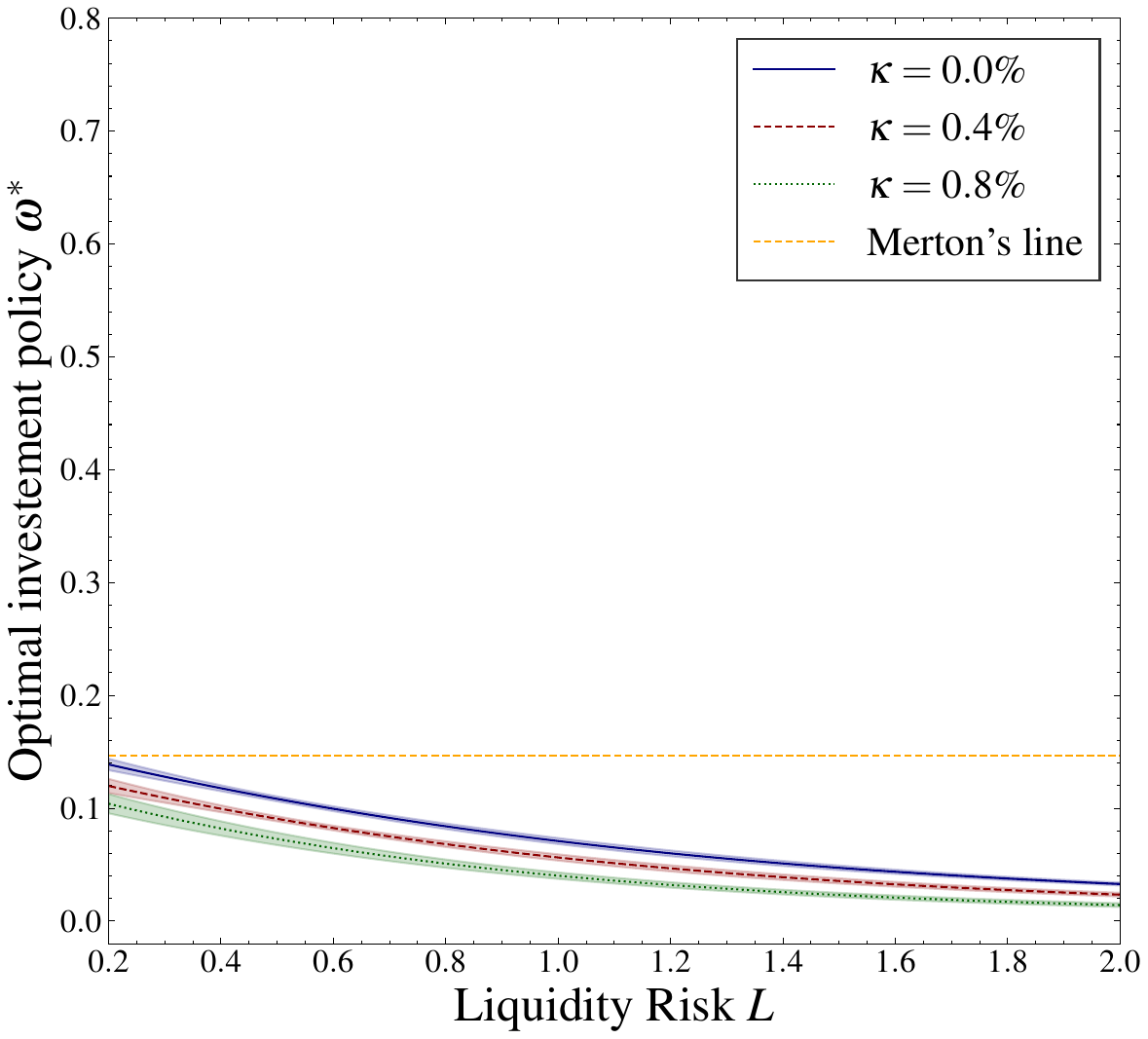}
        \caption{Different $\kappa$}
        \label{fig:exp-L-kappa}
    \end{subfigure}
    \caption{The variation with liquidity risk for $W=2.5$ and $t=0.5$.}
    \label{fig:exp-L}
\end{figure}

\begin{figure}[H]
    \centering
    \begin{subfigure}[b]{0.3\textwidth}
        \centering
        \includegraphics[width=\textwidth]{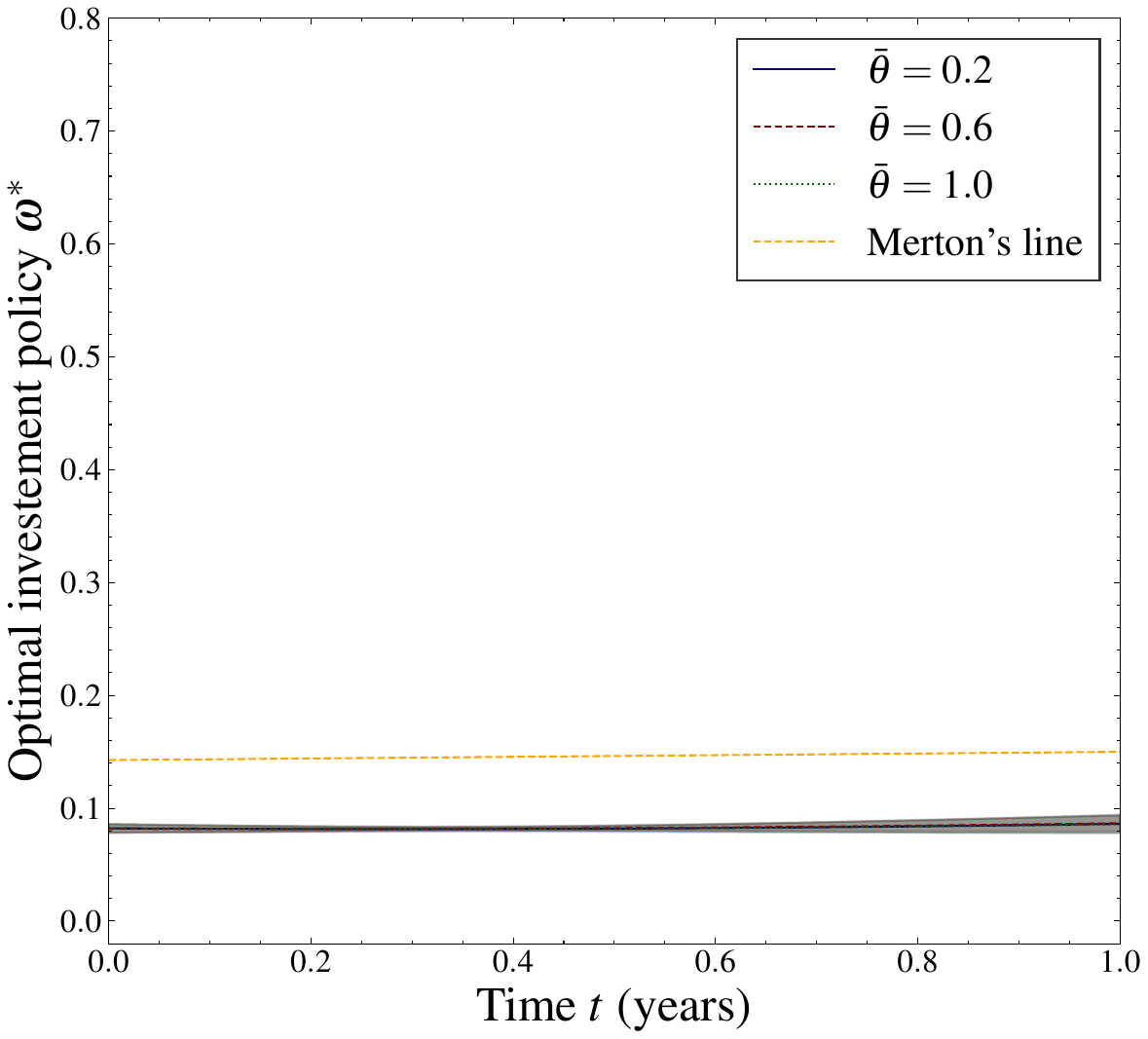}
        \label{fig:exp-L-theta-1}
    \end{subfigure}
    \hspace{1em}
    \begin{subfigure}[b]{0.3\textwidth}
        \centering
        \includegraphics[width=\textwidth]{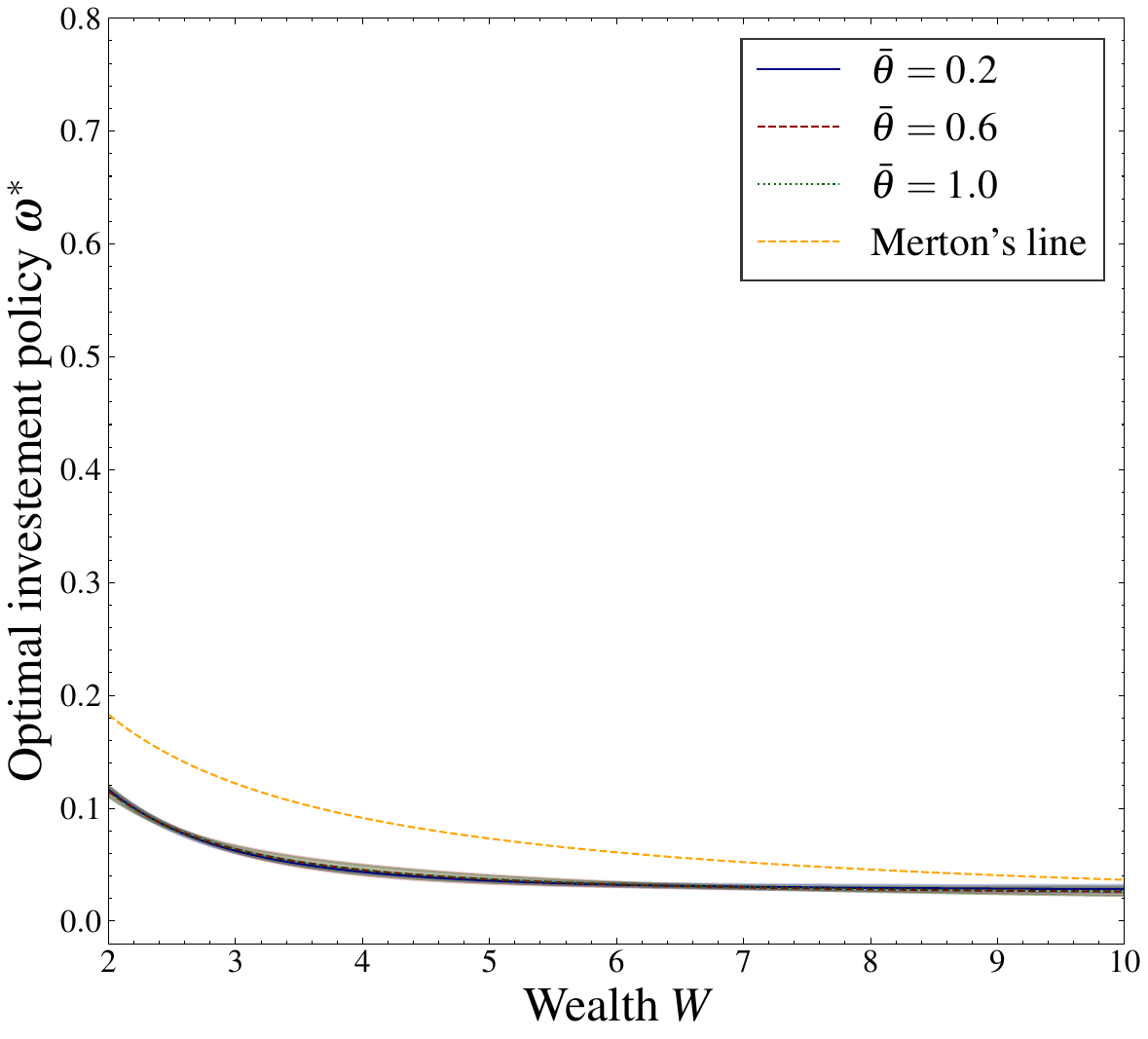}
        \label{fig:exp-L-theta-2}
    \end{subfigure}
    \hspace{1em}
    \begin{subfigure}[b]{0.3\textwidth}
        \centering
        \includegraphics[width=\textwidth]{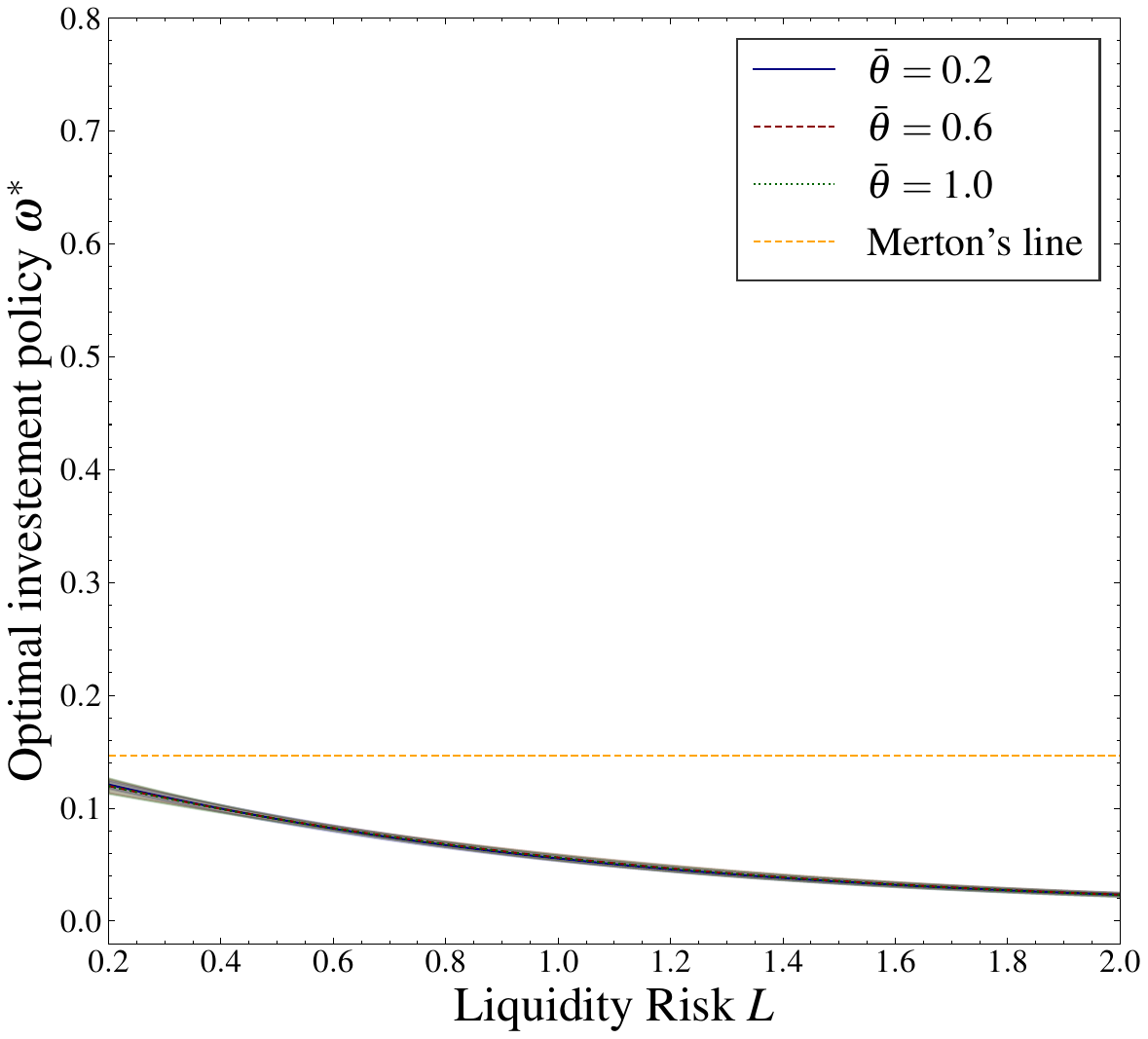}
        \label{fig:exp-L-theta-3}
    \end{subfigure}
    \caption{The variation with $\bar{\theta}$ for $W=2.5$, $L=0.6$ and $t=0.5$.}
    \label{fig:exp-L-theta}
\end{figure}
Under the exponential utility, the optimal policy is dependent on time, wealth, and liquidity risk when both exogenous and endogenous transaction costs are taken into consideration. As shown in Figure \ref{fig:exp-W}, with fixed values of $\beta$ and $\kappa$, when an investor's wealth is small, the effects of these two parameters on investment decisions are evident. Conversely, when an investor's wealth is large enough, these effects can be ignored because the proportion of total costs relative to wealth becomes negligible. 

From these three examples with CRRA and CARA utility functions, one can observe that transaction costs affect an investor's portfolio choice, especially for individual investors. The effects of different types of transaction costs, i.e., exogenous and endogenous transaction costs, on portfolio selection or the pricing of financial derivatives should be analyzed thoroughly.

\section{Conclusion}
In this paper, we present a full study of the portfolio selection problem, considering both exogenous and endogenous transaction costs within the framework of utility maximization theory. Exogenous transaction costs are defined as proportional transaction costs, whereas endogenous transaction costs arise from liquidity risk, which is characterized by a stochastic process. Additionally, we take into account the intrinsic connection between exogenous and endogenous transaction costs. To effectively solve the associated nonlinear two-dimensional HJB equation, we propose an innovative deep learning-driven policy iteration scheme that potentially addresses the curse of dimensionality, which often poses challenges in financial modeling. We also discuss the numerical analysis of the proposed scheme, including convergence analysis in a general setting. Through numerical experiments involving three examples with different utility functions, our findings suggest that traditional portfolio strategies should be adapted to account for both exogenous and endogenous transaction costs.

\begin{appendices}
\section{Components of neural networks}
\label{appendixA}
For each layer, the components for the values network $Q_\phi=f_2^Q \circ \mathbf{tanh} \circ f_1^Q$ are defined as
\begin{align*}
f_2^Q  &: \R^{N^Q} \to \R, \quad x \mapsto W_2^Q x + b_2^Q, \\
\mathbf{tanh}  &: \R^{N^Q} \to \R^{N^V}, \quad x \mapsto \tanh(x), \\
f_1^Q  &: \R^{3} \to \R^{N^Q}, \quad x \mapsto W_1^Q x + b_1^Q,
\end{align*}
and the components for the control network $\omega_\psi=\text{sigmoid} \circ f_2^\omega \circ \mathbf{tanh} \circ f_1^\omega$ are given by
\begin{align*}
\mathrm{sigmoid} &: \R \to \R, \quad x \mapsto \frac{1}{1 + \mathrm{e}^{-x}}, \\
f_2^\omega  &: \R^{N^\omega} \to \R, \quad x \mapsto W_2^\omega x + b_2^\omega, \\
\mathbf{tanh}  &: \R^{N^\omega} \to \R^{N^\omega}, \quad x \mapsto \tanh(x), \\
f_1^\omega &: \R^{3} \to \R^{N^\omega}, \quad x \mapsto W_1^\omega x + b_1^\omega.
\end{align*}
Here, $N_Q$  and  $N_\omega$ denote the hidden layer sizes for each network. Unless specified otherwise, these values are set to 128. Note that both networks are optimized using the Adam optimizer \cite{Kingma2014}, with a learning rate set to $10^{-3}$.

\end{appendices}

\bibliographystyle{abbrv}
\bibliography{ref}

\end{document}